\documentclass[preprint2]{aastex}

\newcommand{\kms}{{km~s$^{-1}$}}

\shorttitle{Opaque Atomic Gas and Star Formation Density in M31}
\shortauthors{Braun et al.}

\begin{document}


\title{A Wide-field High Resolution \ion{H}{1} Mosaic of Messier 31 \\
    I. Opaque Atomic Gas and Star Formation Rate Density}


\author{R. Braun}
\affil{CSIRO - ATNF, PO Box 76, Epping, NSW 1710, Australia}
\email{Robert.Braun@csiro.au}

\author{D.A. Thilker}
\affil{Center for Astrophysical Sciences, Johns Hopkins
	      University, 3400 North Charles Street, Baltimore, MD
	      21218, USA}
\author{R.A.M. Walterbos}
\affil{Department of Astronomy, New Mexico State University,
	      P.O. Box 30001, MSC 4500, Las Cruces, NM 88003, USA}

\and

\author{E. Corbelli}
\affil{INAF - Osservatorio Astrofisico di Arcetri, Largo
	      E. Fermi 5, 50125 Firenze, Italy}




\begin{abstract}
We have undertaken a deep, wide-field \ion{H}{1} imaging survey of
 M31, reaching a maximum resolution of about 50~pc and 2~\kms\ across
 a 95$\times$48~kpc region. The \ion{H}{1} mass and brightness
 sensitivity at 100~pc resolution for a 25~\kms\ wide spectral feature
 is 1500~M$_\odot$ and 0.28~K. Our study reveals ubiquitous \ion{H}{1}
 self-opacity features, discernible in the first instance as
 filamentary local minima in images of the peak \ion{H}{1} brightness
 temperature. Local minima are organized into complexes of more than
 kpc length and are particularly associated with the leading edge of
 spiral arm features. Just as in the Galaxy, there is only patchy
 correspondence of self-opaque features with CO(1-0) emission. We have
 produced images of the best-fit physical parameters; spin
 temperature, opacity-corrected column density and non-thermal
 velocity dispersion, for the brightest spectral feature along each
 line-of-sight in the M31 disk. Spectroscopically opaque atomic gas is
 organized into filamentary complexes and isolated clouds down to
 100~pc. Localized opacity corrections to the column density exceed an
 order of magnitude in many cases and add globally to a 30\% increase
 in the atomic gas mass over that inferred from the integrated
 brightness under the usual assumption of negligible
 self-opacity. Opaque atomic gas first increases from 20 to 60~K in
 spin temperature with radius to 12~kpc but then declines again to
 20~K beyond 25~kpc. We have extended the resolved star formation law
 down to physical scales more than an order of magnitude smaller in
 area and mass than has been possible previously. The relation between
 total-gas-mass- and star-formation-rate-density is significantly
 tighter than that with molecular-mass and is fully consistent {\it in
 both slope and normalization\ } with the power law index of 1.56 found in the
 molecule-dominated disk of M51 at 500~pc resolution.  Below a
 gas-mass-density of about 5 M$_\odot$pc$^{-2}$, there is a down-turn
 in star-formation-rate-density which may represent a real local
 threshold for massive star formation at a cloud mass of about 5$\cdot
 10^4$~M$_\odot$.

\end{abstract}


\keywords{galaxies: individual (M31) -- galaxies: Local Group --
   galaxies: ISM --  radio lines: galaxies -- stars: formation}



\section{Introduction}
As the nearest major external galaxy and possibly the dominant member
of the Local Group, Messier~31 has received a great deal of attention
since its' first documented telescopic observation by Simon Marius in
1612 \citep{mari12}. The many stellar and gaseous constituents of M31
continue to be studied with ever greater sensitivity and resolution,
permitting insights that are in many cases difficult or even
impossible to achieve for our own Galaxy given our vantage point
within the disk. One of these areas, where an external yet nearby
vantage point is crucial to achieving a global overview, is the study
of the neutral interstellar medium. Although we can achieve remarkable
physical sensitivity and resolution of the neutral medium of the
Galaxy, resolving individual structures of 100's of AU in size and as
little as several Jupiter masses \citep{brau05}, it is very difficult
to estimate even the total mass of the neutral medium, let alone its
overall distribution and relationship with other components. Studies
of the neutral medium of M31 began with the pioneering work of
\citet{huls57} as one of the first programs undertaken with the
Dwingeloo 25m telescope. Subsequent studies with larger single dishes
and then with interferometric arrays have provided enhanced resolution
and brightness sensitivity, although the highest resolution studies
have often only been directed at a portion of the disk
(eg. \citet{brin84}, \citet{brau91}, \citet{cari06}).

In this paper we describe our Westerbork and Green Bank Telescope
program to image the entire atomic disk of M31 with a uniformly high
sensitivity and resolution. The most recent effort on a comparable
scale was that of \citet{unwi83} which achieved arcmin spatial and 16
\kms\ velocity resolution with an {\sc RMS} noise of 3.6~K in two
elliptical fields covering about 12 square degrees. By contrast we
will present imaging down to 15'' spatial and 2.3 \kms\ velocity
resolution (sampled at 2 \kms) with a uniform 2.7~K {\sc RMS} over a
region of about 24 (7$\times$3.5) square degrees. This corresponds to
a maximum linear resolution of about 50~pc over a total extent of more
than 90~kpc. With even modest spatial smoothing we reach sub-Kelvin
brightness sensitivity (for example 0.37~K at 1 arcmin), making our
database the most detailed documentation yet of the neutral medium of
any complete galaxy disk, including our own. For comparison, the
Leiden/Argentine/Bonn Survey of Galactic \ion{H}{1} has achieved 36
arcmin and 1.3~\kms\ resolution over 360 degrees and reaches 0.08~K
{\sc RMS} \citet{kalb05} and the Large Magellanic Cloud survey of
\citet{kim03} achieves 1~arcmin and 1.7~\kms\ resolution over
12$\times$10 degrees and reaches 2.5~K {\sc RMS}.

We begin with a description of the observations and their reduction in
\S\ref{sec:obse} and continue with an overview of the results in
\S\ref{sec:resu}. In \S\ref{sec:disc} we present some initial analysis
of the data, beginning with a discussion of \ion{H}{1} self-opaque
features and followed by a detailed fitting of all high
signal-to-noise spectra to determine the best fitting combination of
the physical parameters along each line-of-sight: cool component
temperature, total column density and non-thermal velocity
dispersion. We continue with the analysis of radial dependence for the
relevant physical parameters and finally we focus on the relationship
between surface density of gas mass and star formation rate.  We close
with some brief conclusions in \S\ref{sec:conc}. We assume a distance
to M31 of 785~kpc \citep{mcco05} throughout.

Further papers in this series will include: (1) a detailed
kinematic- and mass-model of M31, (2) comparative studies of
the morphology of the inner and outer disks and (3) a study of small-scale
kinematic features such as outflows and other peculiar velocity components.


\section{Observations and Reduction}
\label{sec:obse}
Interferometric observations of the extended M31 \ion{H}{1} disk with
the Westerbork Synthesis Radio Telescope (WSRT) array were obtained on
a Nyquist sampled (15 arcmin spacing) pointing grid defined in the M31
(major, minor) axis coordinate system. The disk extent was first
determined in a precursor total power survey which fully sampled a
6$\times$6 degree field at low resolution (36' FWHM) using the 14
individual telescopes of the WSRT array. A total of 163 synthesis
pointing positions were chosen from the grid to provide good coverage
of the region of anticipated \ion{H}{1} emission. Observations were
then obtained for 27 tracks of 12 hour duration with the WSRT array,
each sampling 6 (or in one case 7) different pointing positions. These
data were observed between August 2001 and January 2002. Each 12 hour
track involved cycling through the 6 relevant pointing positions with
a series of 10 minute snap-shots. The tracks were preceeded by half
hour observations of the calibration sources 3C286 and CTD93 and
followed by half hour observations of 3C147 and CTA21. A total
band-width of 5~MHz (covering heliocentric radial velocities between
$-$828 and $+$227~\kms) was observed with 512 spectral channels in two
linear polarization products yielding a channel width of
2.06~\kms\ and resolution of 2.27~\kms. Calibration and flagging of the data
were done in Classic AIPS. After external band-pass and gain
calibration, each pointing was self-calibrated using the continuum
emission which happened to be present. The continuum model was
subtracted from the visibilities and residual continuum was removed
with UVLIN.

Sensitive total power data covering a 7$\times$7 degree field was
acquired with the Green Bank Telescope (GBT) telescope over six nights
in September 2002. These data, with 9\farcm1 FWHM $\times$
1.03~\kms\ resolution and an RMS sensitivity in \ion{H}{1} column
density over relevant linewidths of about 5$\times10^{17}$~cm$^{-2}$,
have already been presented in \citet{thil04}.

A joint deconvolution of all 163 pointings was done with the Miriad
\citep{saul95} task MOSSDI, after forming a weighted sum of the
linearly combined synthesis image (employing uniform weighting
followed by a Gaussian taper) with the relevant GBT total power
image. Each of the 163 synthesized beams was also modified by addition
of a Gaussian approximation of the total power beam using the same
relative weightings as used when summing the synthesis and total power
images. The relative weightings were determined empirically by
minimizing the ``short-spacing bowl'' effect in the synthesis beams,
but they correspond roughly to the inverse of the relative beam
areas. The exact choice of weighting is not critical, in theory, since
the process of deconvolution should yield an ``ideal'' final PSF if
all emission has been successfully incorporated into the component
model. In practise it is never possible to incorporate all emission
into the component model, so it is important to make the ``dirty'' PSF
as clean as possible (particularly with respect to diffuse emission)
by an optimized choice of weights.

The linearly combined ``dirty'' mosaic images were multiplied with the
combined WSRT survey sensitivity pattern in order to weight down the
noise around the periphery of the sampled region before
deconvolution. If this were not done, then the noisy periphery would
be severely over-CLEANed relative to the central regions, yielding
greatly reduced image fidelity. The assumption underlying this spatial
tapering is that the region of significant emission is contained
within the sampling pattern. We will comment below on the extent to
which this assumption was met in practise. The 300 semi-independent
spectral channels (each of 2~\kms\ width) containing M31's \ion{H}{1}
emission (heliocentric velocities of about $-$621 to $-$20 \kms) were
processed individually. Foreground \ion{H}{1} emission from the Galaxy
is present in the field of M31 at velocities between about $-$130 and
$+$45 \kms. Reduction of the WSRT data over the full velocity range of
Galactic emission was not undertaken at this time. The GBT total power
images at velocities between $-$130 and $-$20 \kms\ were modified by
subtraction of the median value determined outside of the region
occupied by the M31 disk before combination with the WSRT mosaic to
simplify the deconvolution problem. A positive pedestal of emission
underlying the entire galaxy is a needless complication that 
undermines image fidelity in the deconvolved result. 

Given the non-linear nature of the deconvolution process and the
desire to probe a variety of different output spatial
(``full''=18\farcs4$\times$14\farcs8 (NS$\times$EW), 30\arcsec,
60\arcsec and 120\arcsec) and velocity (2.3, 6.2, 10.3, and 18.5 \kms)
resolutions, we carried out the joint deconvolution independently for
each different combination of these parameters. Deconvolution was in
all cases done to a depth of twice the theoretical {\sc RMS} noise
level at each resolution. A bug in the MOSSDI source code was remedied
(affording it protection against a negative loop gain). A small
adjustment was also made to the primary beam model employed within
Miriad for the WSRT telescopes. It was found empirically, that a small
increase (by 5 \%) in the assumed FWHM of the analytic approximation
($cos^6(C \cdot \nu \cdot r)$ for a constant $C$ at frequency $\nu$ in
terms of the radius $r$) for the beam pattern yielded the highest
dynamic range in the mosaic result. The slightly larger than expected
beam size (2110 arcsec FWHM rather than 2013 arcsec at 1420 MHz) has
since been verified by direct measurement \citep{popp07}. The
resulting {\sc RMS} noise levels at various spatial and velocity
resolutions are listed in Table~\ref{tab:sens}. Note that unlike the
case of total power observations of Galactic \ion{H}{1}, the detected
M31 \ion{H}{1} emission in our synthesis images does not contribute
significantly to the measured antenna temperature of each dish, and
therefore does not give rise to an increase of the system temperature.

\begin{deluxetable}{ccccccc}
\tablecolumns{7}
\tablewidth{0pt}
\tablecaption{RMS Sensitivity at Various Resolutions}
\tablehead{
\multicolumn{2}{c}{Resolution} & \colhead{} &
\multicolumn{4}{c}{RMS Sensitivity} \\
\cline{1-2} \cline{4-7} \\
\colhead{Beam} & \colhead{$\Delta V$} & \colhead{} & \colhead{$\Delta S$} & 
\colhead{$\Delta T_B$} & \colhead{$\Delta N_{HI}$}& \colhead{$\Delta M_{HI}$} \\
\colhead{[arcsec]} & \colhead{[km s$^{-1}$]} & \colhead{} &\colhead{[mJy/Beam]} &
\colhead{[K]} & \colhead{[10$^{18}$cm$^{-2}$]} &
\colhead{[M$_\odot$]}
} 
\startdata
18$\times$15 & 2.3 &\  & 1.2 & 2.7 & 11.3 & 390 \\
30 & 2.3 &\ & 1.5 & 1.0 & 4.2 & 490 \\
60 & 2.3 &\ & 2.2 & 0.37 & 1.6 & 720 \\
120 & 2.3 &\ & 4.2 & 0.15 & 0.63 & 1400 \\
 & & & & & & \\
18$\times$15 & 18.5 &\ & 0.42 & 0.95 & 32 & 1100 \\
30 & 18.5 &\ & 0.53 & 0.35 & 11.8 & 1400 \\
60 & 18.5 &\ & 0.78 & 0.13 & 4.4 & 2100 \\
120 & 18.5 &\ & 1.5 & 0.053 & 1.8 & 4000 \\
\enddata
\label{tab:sens}
\end{deluxetable}

A locally modified version of Miriad was employed that contains many
enhancements for high performance computing. It extensively exploits
parallelism at several levels: multi-threading (on machines that have
multiple CPUs) and vectorization (using SSE (Streaming SIMD
Extensions) instructions which are available on most contemporary
processors).  Furthermore, we used task-level parallelism by dividing
different reduction pipelines (specific resolutions and spectral
channels) over multiple machines. The ``Netherlands Grid''
infrastructure provided us with the resources to do the computational
work. Using these techniques, we were able to perform all reductions
(16 different resolutions, 300 semi-independent spectral channels) in
36 hours of elapsed time. A previous attempt to carry out the same
processing, before code optimization and using only a single high-end
server, required several months of elapsed time.

Although data products at all 16 different resolutions were generated
directly from the visibility data, we also produced velocity-averaged
data products at a fixed spatial resolution by averaging the
individually processed channels after deconvolution. Superior dynamic
range was found for these post-deconvolution averaged products at
spatial resolution of both 60 and 120 arcseconds, while the directly
imaged products were superior at full and 30 arcsecond
resolutions. Apparently, different deconvolution limitations dominate
in these two regimes. At high spatial resolutions, the increased
signal-to-noise ratio afforded by visibility-based velocity smoothing
improves the recovery of extended emission structures, while at low
spatial resolutions we already encounter the finite image fidelity of
the deconvolution in an individual velocity channel. In this case,
averaging spectral images after deconvolution can improve the dynamic
range of the result. The nominal {\sc RMS} noise levels after spectral
smoothing scale simply with the square root of the eventual velocity
resolution. \ion{H}{1} column density sensitivity at an appropriate
spectral smoothing for faint emission features (of 18.5~\kms) is
tabulated in Table~\ref{tab:sens} at a series of spatial resolutions.

Images of the peak observed line brightness and the integrated
brightness and associated mean velocity (the 0$^{th}$ and 1$^{st}$
moments) were produced at each resolution after first masking the
data-cube. The mask was constructed from the cube of 120 arcsec and 18
\kms\ resolution, by first blanking brightnesses below 1.5$\sigma$ and
then interactively masking with an arbitrary polygon the ``island'' of
most significant emission at each velocity. Interactive isolation of
M31 \ion{H}{1} emission was most critical in the velocity range that
overlaps that of the Galaxy ($-$130 to $-$20 \kms). Subtraction of the
off-M31 median total power before deconvolution already served to
place the M31 emission on a plausible base level. Only a single
contiguous (both spatially and in velocity) ``island'' of emission at
each velocity was included in the blanking mask. Although
isolated features in the field may be associated with M31, we are
unaware of how to distinguish them from features in the Galactic
foreground. All features in the field which were not
contiguously connected with the M31 disk emission were presumed to be
part of the Galactic foreground and were excluded.

Images of integrated brightness were rescaled to represent column
density assuming a negligible \ion{H}{1} line opacity. Although this
is the usual assumption, it is clear that it is incorrect in
practise. \citet{brau92} have applied a simple,
single temperature opacity correction to the observed \ion{H}{1}
brightnesses from the North-East half of M31 and find local
corrections of a factor of 2 and an integrated correction of some
20\%. We address the issue of \ion{H}{1} opacity in greater detail
below by directly fitting physical models to individual spectra for
all lines-of-sight.

Another useful type of image for diagnostic purposes is that of
``velocity coherence'' (VC) (as first introduced by \citet{brau97}),
defined as the ratio of peak brightness at two different spectral
resolutions. For example,
VC$_{30/5}(x,y)~=~P_{30}(x,y)$ / $P(_{5}x,y)$, where
$P_{R}(x,y)$ is the peak brightness observed at a spectral
resolution of $R$~km~s$^{-1}$. In the absence of noise, $0 \le VC \le
1$, when comparing coarser with finer spectral resolutions, since an
infinitely broad spectral feature will always retain the same peak
brightness (VC~=~1), while an extremely narrow feature will be
strongly diluted (VC~$\rightarrow$~0). Images of
velocity coherence are a robust measure of linewidth, since no
assumption regarding the intrinsic line-shape, nor reliance on a
fitting algorithm are required for it's calculation. Moreover, even
when spectra become multi-valued, as they do in highly inclined (like
M31) or kinematically disturbed systems, the VC can continue to probe
the attributes of the dominant spectral feature along each
line-of-sight rather than a smeared version of all spectral
content. In the particular case of an intrinsically Gaussian line
profile with dispersion, $\sigma$, that is fully resolved at the
higher resolution and then convolved with a Heaviside (top-hat)
function of width, $\Delta V$, one gets the relation, $\sigma = \Delta
V \cdot VC/[(1-VC)\sqrt{2\pi}]$.

\section{Overview of Results}
\label{sec:resu}

\begin{figure*}
   \centering
   \includegraphics[width=16.5cm]{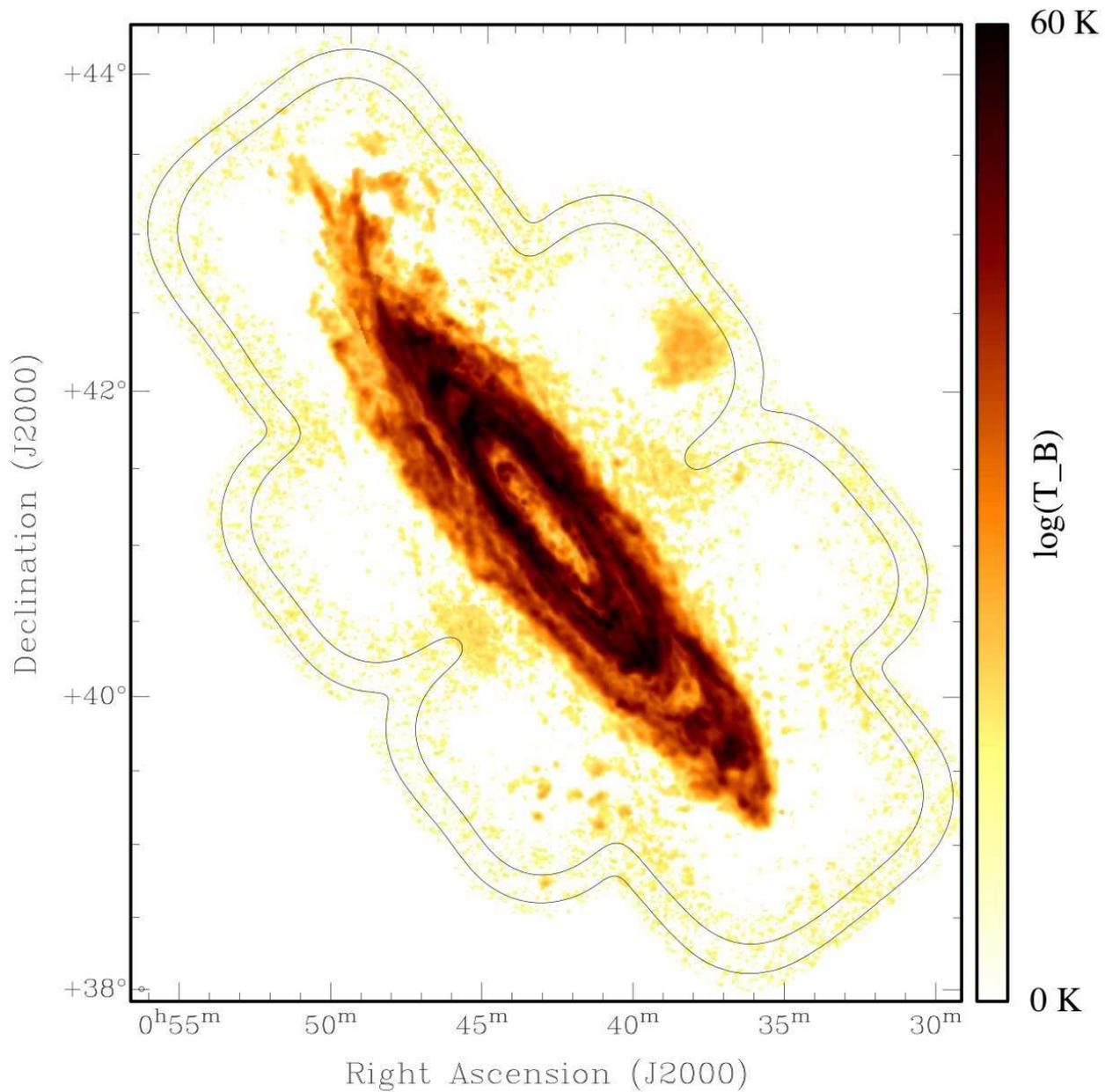}
      \caption{Peak brightness of \ion{H}{1} emission determined at
      120 arcsec and 10 \kms\ resolution is overlaid with the combined
      sensitivity pattern of the 163 pointings in the mosaic
      observation. Contours are drawn at 10 and 50\% of the peak
      sensitivity. Peak brightness is shown on a logarithmic scale
      which saturates at 60~K. The beam FWHM is indicated in the lower
      left corner.}
         \label{fig:mp120overfig}
   \end{figure*}

An overview of the joint deconvolution result is shown in
Figure~\ref{fig:mp120overfig} where the WSRT mosaic sensitivity
pattern is overlaid on a low resolution (120 arcsec, 10~\kms) image of
peak \ion{H}{1} brightness. The Karma package \citep{gooc96} has
been employed for many of the illustrations. Sensitivity contours are
shown at 10 and 50\% of the peak sensitivity which results from the
sampling pattern. The combined sensitivity pattern has what turn out
to be unfortunate local minima on either side of the minor
axis. Although little evidence for emission was discerned at these
locations in our pilot WSRT total power observations, it is clearly
present in the more sensitive total power data that we have acquired
since (\citet{thil04} and \citet{brau04}). Diffuse patches of emission
are apparent on both sides of the minor axis. The apparent termination
of these minor axis extensions is likely influenced by the
multiplication of the brightness distribution with the sensitivity
pattern (to control noise amplification during deconvolution, as
outlined in \S\ref{sec:obse} above). More extensive sampling of the
minor axis with additional mosaic pointings will be required for high
fidelity imaging at arcmin resolution. Our total power data (with up
to 9 arcmin resolution) do not suffer from this limitation.  Similar
attenuation affects Davies' Cloud (\citet{davi75}, \citet{heij02}) the
diffuse structure at $\delta$=42\fdg5 to the northwest of the M31
disk. The southernmost of the \ion{H}{1} clumps at $\delta$=38\fdg8
and due South of M31 \citep{thil04} is also somewhat
attenuated. Sensitive pointed observations of these features are
presented in \citet{west05}.

\begin{figure*}
   \centering
   \includegraphics[width=16.5cm]{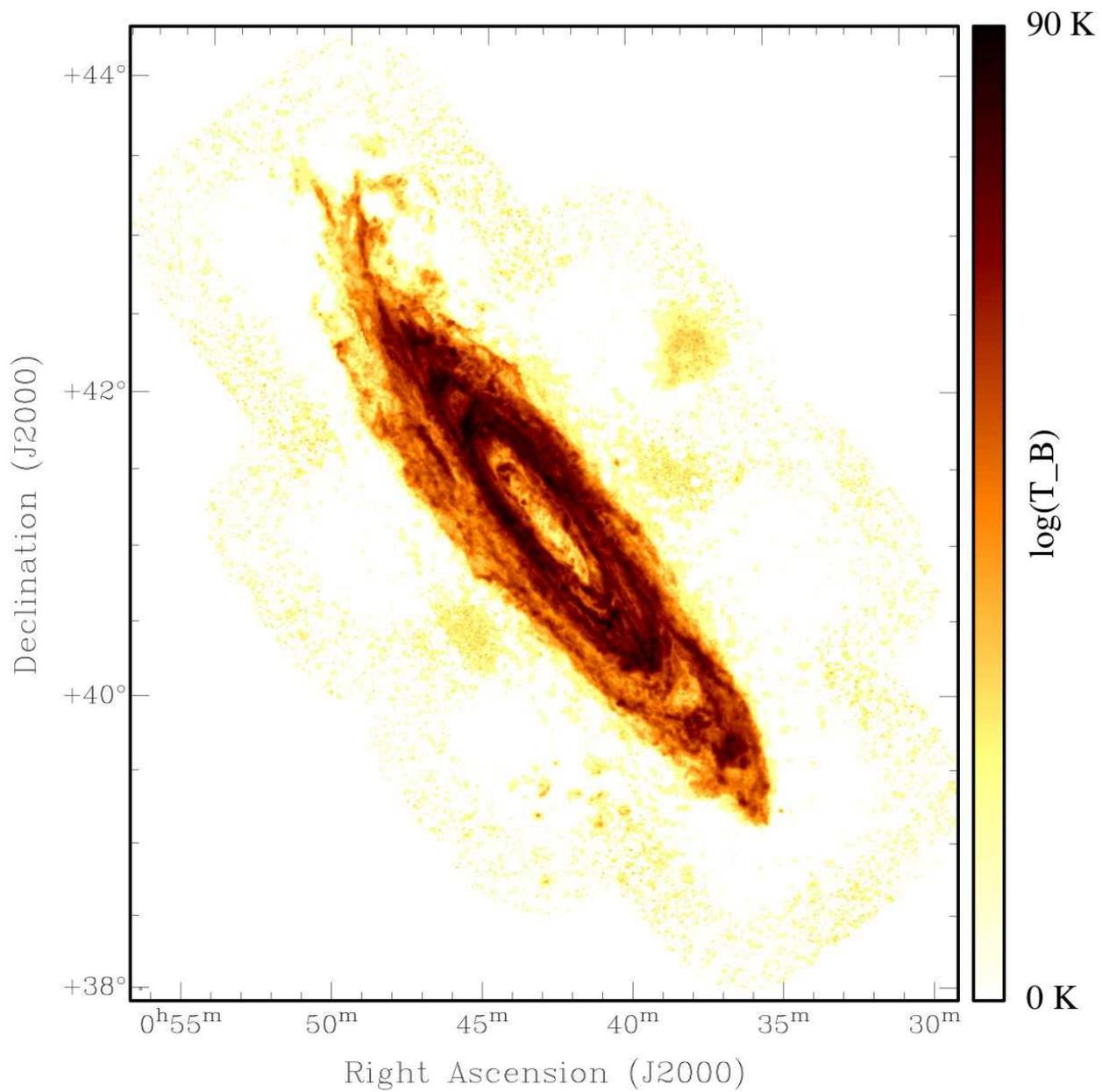}
      \caption{Peak brightness of \ion{H}{1} emission determined at 60
      arcsec and 6 \kms\ resolution. Peak brightness is shown on a
      logarithmic scale which saturates at 90~K. The beam FWHM is
      indicated in the lower left corner.}
         \label{fig:mp60fig}
   \end{figure*}

\begin{figure*}
   \centering
   \includegraphics[width=16.5cm]{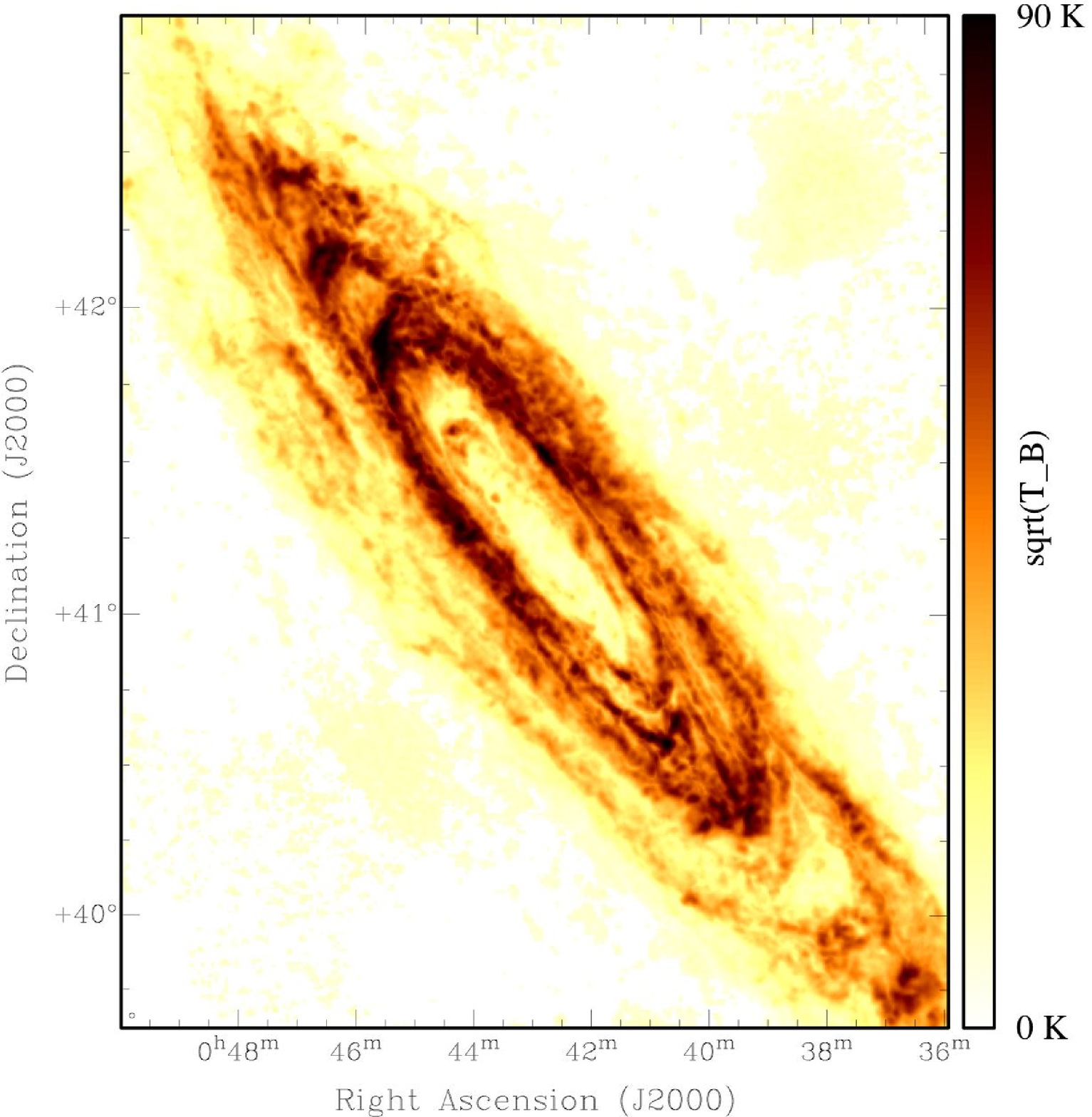}
      \caption{Peak brightness of \ion{H}{1} emission determined at 60
	arcsec and 6 \kms\ resolution of the central 50\% of the survey
	region. Peak brightness is shown on a square-root scale which
	saturates at 90~K. The beam FWHM is indicated in the lower
	left corner.}
         \label{fig:mp60zoom2fig}
   \end{figure*}

\begin{figure*}
   \centering
   \includegraphics[width=16.5cm]{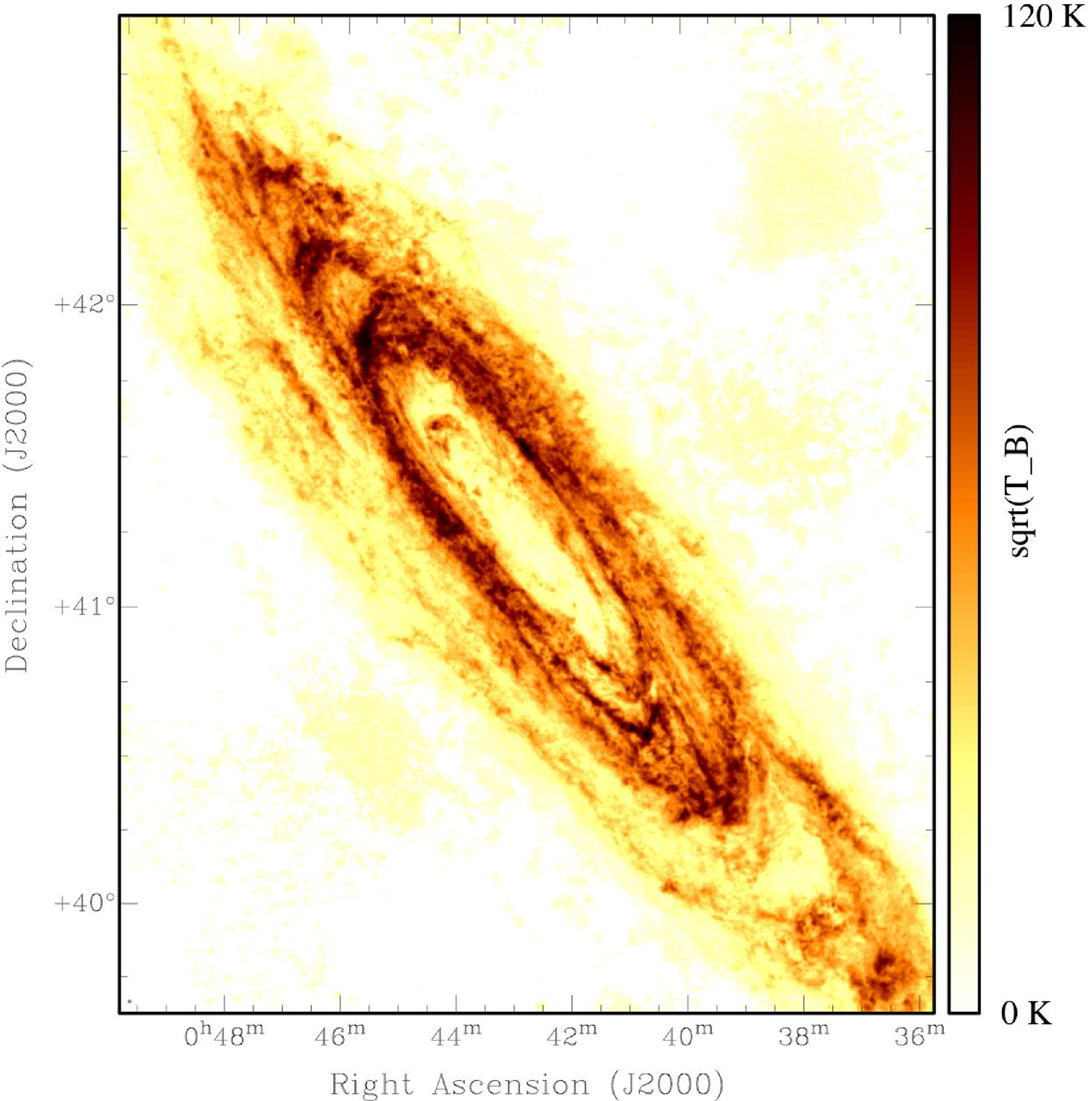}
      \caption{Peak brightness of \ion{H}{1} emission determined at 30
	arcsec and 6 \kms\ resolution of the central 50\% of the survey
	region. Peak brightness is shown on a square-root scale which
	saturates at 120~K. The beam FWHM is indicated in the lower
	left corner.}
         \label{fig:mp30zoom2fig}
   \end{figure*}

\begin{figure*}
   \centering
   \includegraphics[width=16.5cm]{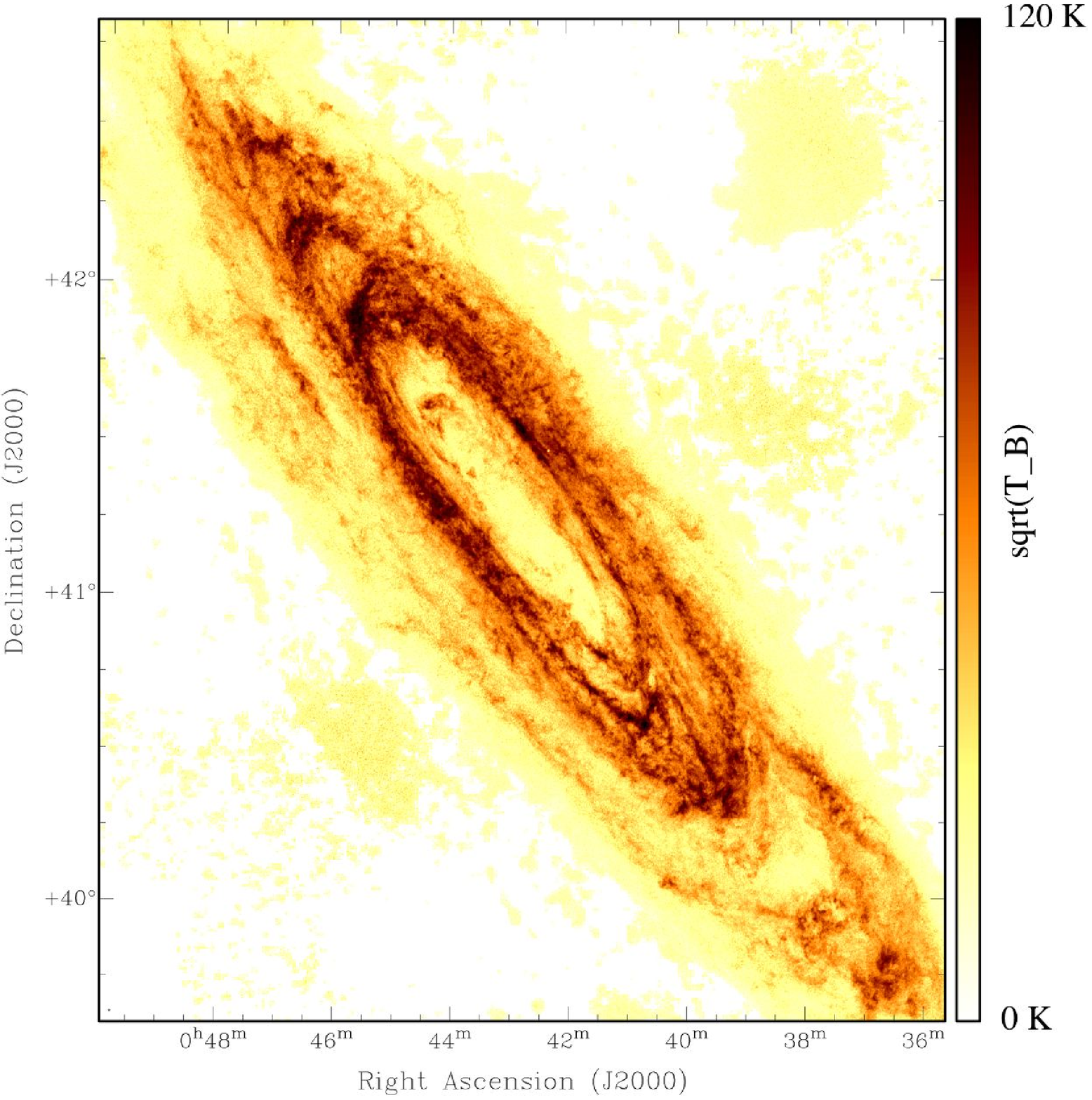}
      \caption{Peak brightness of \ion{H}{1} emission determined at
	about 15
	arcsec and 6 \kms\ resolution of the central 50\% of the survey
	region. Peak brightness is shown on a square-root scale which
	saturates at 120~K. The beam FWHM is indicated in the lower
	left corner.}
         \label{fig:mp15zoom2fig}
   \end{figure*}

Greater detail can be discerned in Figure~\ref{fig:mp60fig} in which
peak brightness is shown as determined at 60 arcsec (220~pc) and 6
\kms\ resolution, although even at this modest resolution, the PSF
(plotted in the lower left corner of the image) is barely resolved on
a printed page, given that the image spans 111~kpc across
the diagonal. Zooming in at this resolution on the central disk as in
Figure~\ref{fig:mp60zoom2fig} finally permits appreciation of which
features are resolved. Jumping now to 30'' (110~pc) resolution on the
same field in Figure~\ref{fig:mp30zoom2fig} begins to reveal the
intricate filamentary structure of the neutral ISM. Point-like local
minima are seen toward the most prominent background sources
(J004218+412926 and J004648+420855) against which \ion{H}{1}
absorption is detected.  Increasing the angular resolution to the
maximum available of about 55~pc (18.4$\times$14.8 arcsec) as in
Figure~\ref{fig:mp15zoom2fig} does not appear to appreciably add
detail to the filamentary structure, while increasing the noise level
significantly. This is consistent with what was found by
\citet{brau95} when analyzing \ion{H}{1} emission data from the
northeast half of the inner M31 disk; beginning with data having 10''
and 5~\kms\ resolution, spatial smoothing to 40'' resolution was
required before the mean peak detected brightness decreased by 8\%,
while velocity smoothing to even 10~\kms\ resulted in a mean decrease
of peak brightness of 15\%. The major \ion{H}{1} cloud complexes which
are discernable in external galaxies with current instrumentation are
essentially resolved with about 100~pc and 5~\kms\
resolution. However, it must also be stressed that significant
structure is present at even the highest angular and velocity
resolution that is available. We will briefly discuss some examples in
\S\ref{sec:disc} below.

\begin{figure*}
   \centering
   \includegraphics[width=16.5cm]{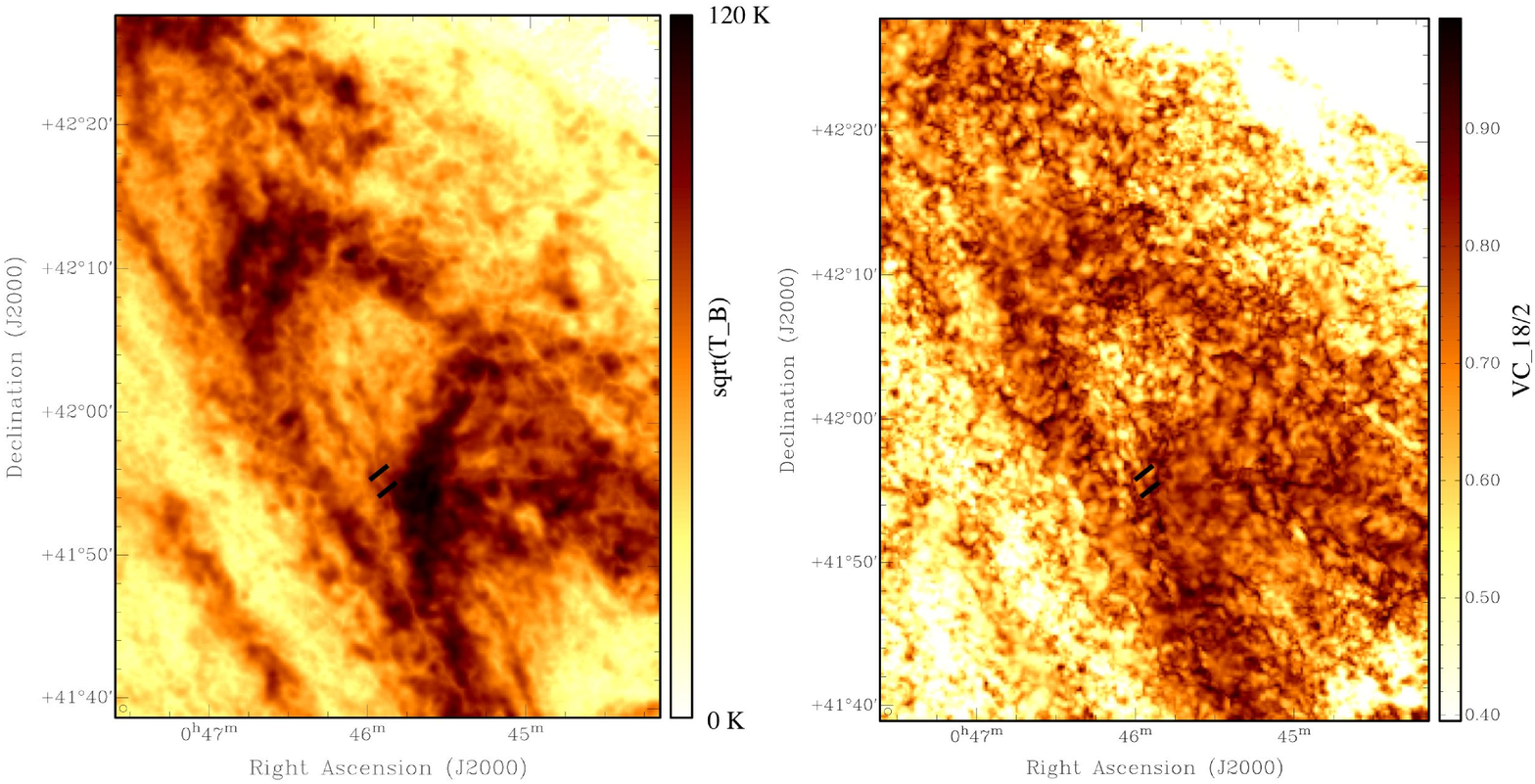} 
      \caption{Peak brightness (left) and velocity coherence (right)
        of \ion{H}{1} emission determined at 30 arcsec resolution of
        the North-East disk region. Peak brightness is shown on a
        square-root scale which saturates at 120~K. Velocity coherence
        (defined in the text) is measured between 18.5 and 2.1
        \kms\ spectral smoothing. The black bars indicate the location
        of the spectra shown in Fig.~\ref{fig:pannefig}. The beam FWHM
        is indicated in the lower left corner. }
         \label{fig:mp30zoomnefig}
   \end{figure*}
\begin{figure*}
   \centering
   \includegraphics[width=16.5cm]{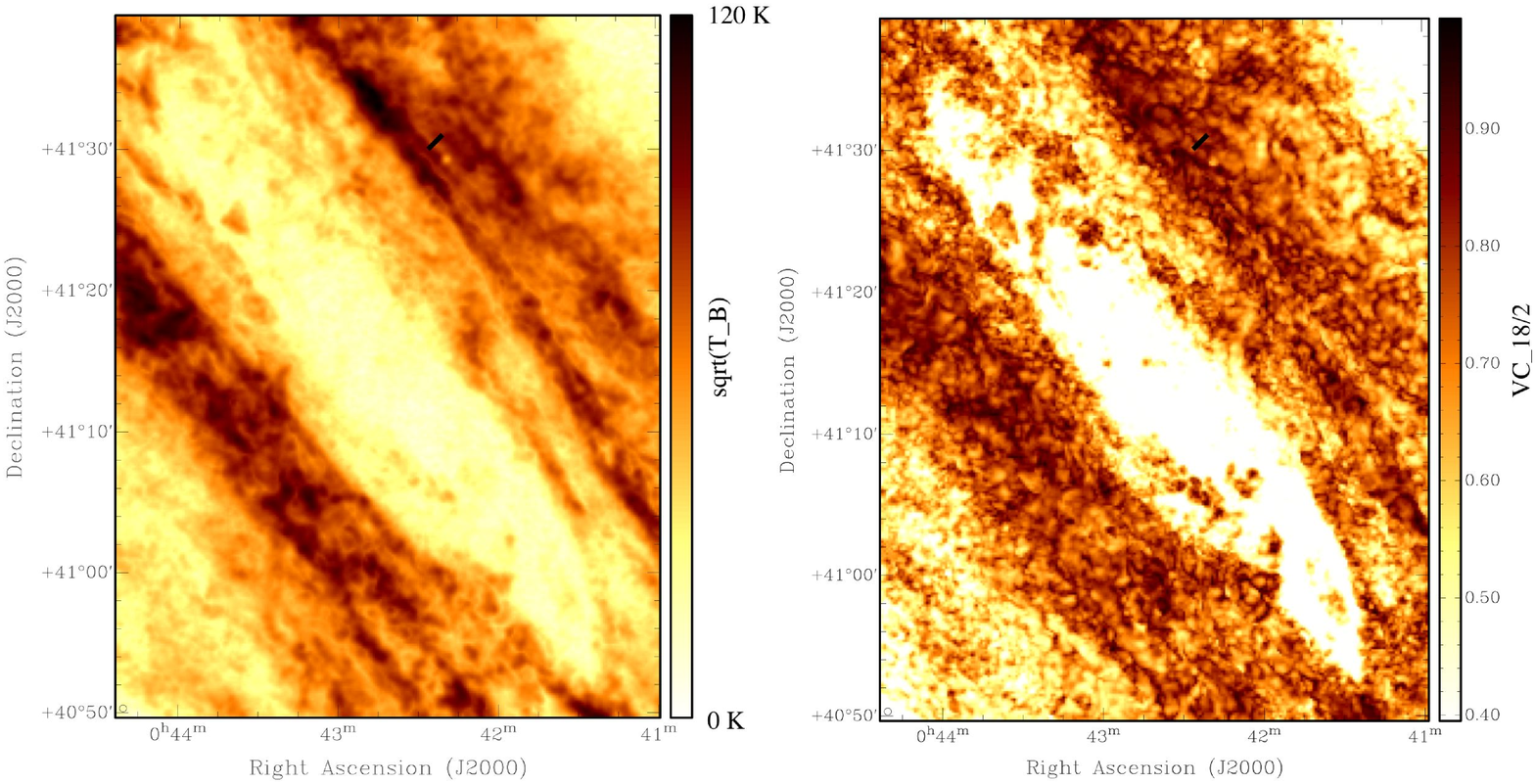} 
      \caption{Peak brightness (left) and velocity coherence (right)
        of \ion{H}{1} emission determined at 30 arcsec resolution of
        the central survey region. Peak brightness is shown on a
        square-root scale which saturates at 120~K. Velocity coherence
        (defined in the text) is measured between 18.5 and 2.1
        \kms\ spectral smoothing. The black bar indicates the location
        of the spectra shown in Fig.~\ref{fig:pannefig}. The beam FWHM
        is indicated in the lower left corner. }
         \label{fig:mp30zoomcfig}
   \end{figure*}
\begin{figure*}
   \centering
   \includegraphics[width=16.5cm]{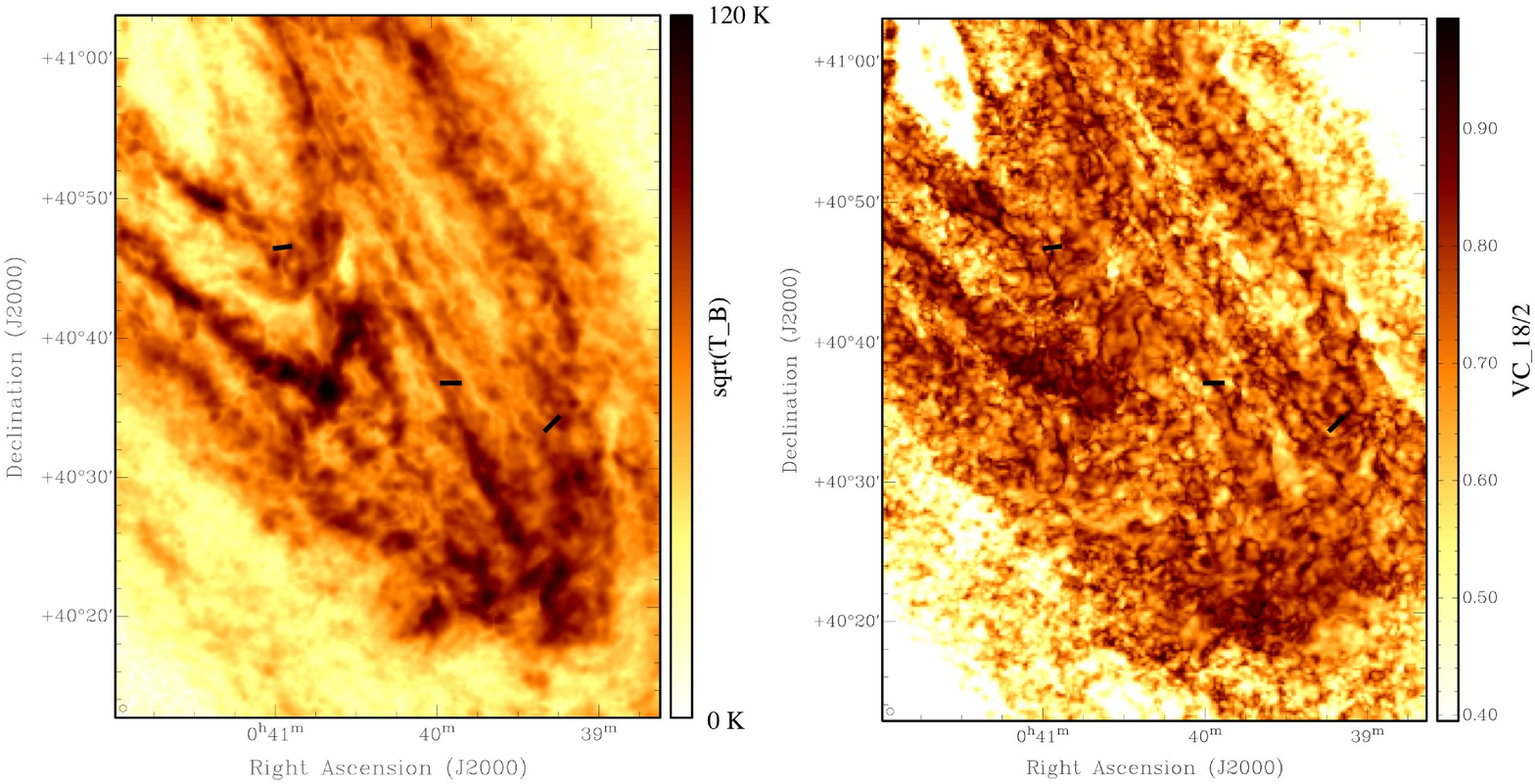} 
      \caption{Peak brightness (left) and velocity coherence (right)
        of \ion{H}{1} emission determined at 30 arcsec resolution of
        the South-West disk region. Peak brightness is shown on a
        square-root scale which saturates at 120~K. Velocity coherence
        (defined in the text) is measured between 18.5 and 2.1
        \kms\ spectral smoothing. The black bars indicate the location
        of the spectra shown in Fig.~\ref{fig:panswfig}. The beam FWHM
        is indicated in the lower left corner. }
         \label{fig:mp30zoomswfig}
   \end{figure*}
\begin{figure*}
   \centering
   \includegraphics[width=16.cm]{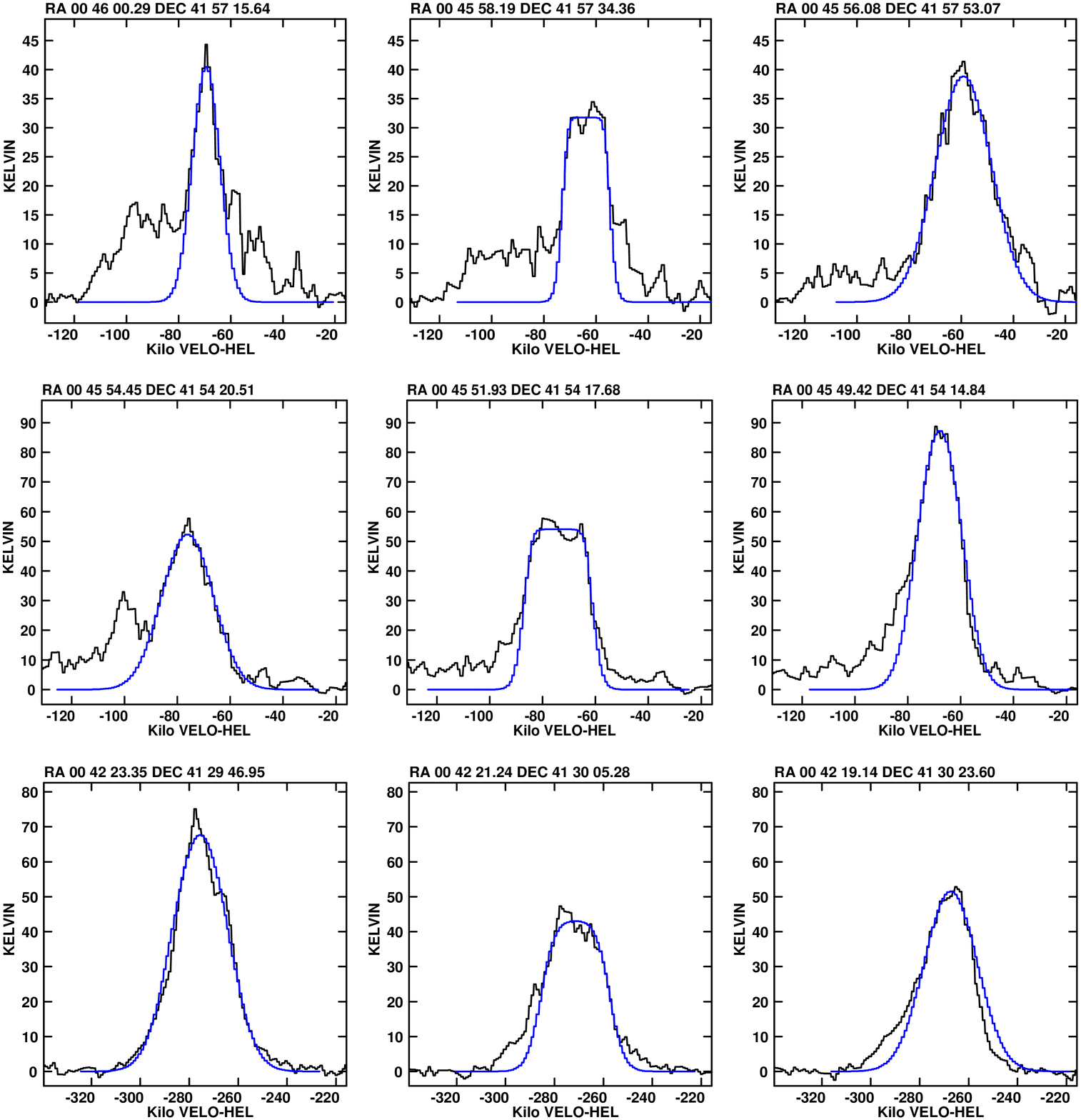}
      \caption{\ion{H}{1} spectra of dark filaments. The three
        sequences of spectra (top, middle and bottom) each represent a
        cross-cut perpendicular to a filamentary local minimum in the
        image of peak brightness temperature shown in
        Figs.~\ref{fig:mp30zoomnefig} and \ref{fig:mp30zoomcfig}. The
        spectrum in each central panel is in the direction of the dark
        filament, while the left and right spectra are offset by about
        30 arcsec (as indicated by the coordinates). The resolution is
        30\arcsec and 2.3~\kms. Note the
        depressed, flat-topped profiles toward the filament, flanked
        by brighter, more smoothly peaked profiles. The smooth curve
        overlaid on each spectrum is the parameterized fit. }
         \label{fig:pannefig}
   \end{figure*}
\begin{figure*}
   \centering
   \includegraphics[width=16.cm]{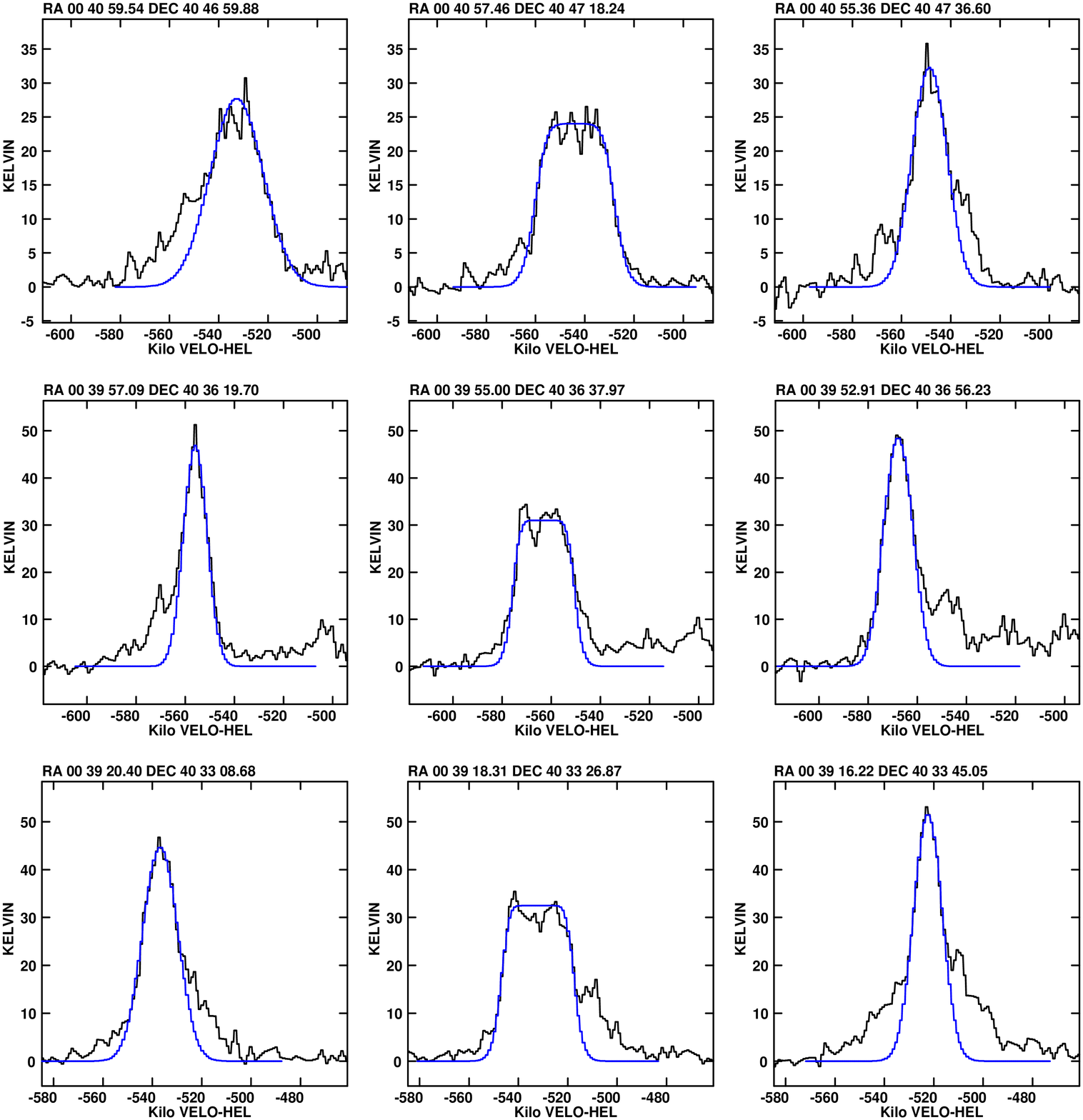}
      \caption{\ion{H}{1} spectra of dark filaments. The three
        sequences of spectra (top, middle and bottom) each represent a
        cross-cut perpendicular to a filamentary local minimum in the
        image of peak brightness temperature shown in
        Fig.~\ref{fig:mp30zoomswfig}. The spectrum in each central
        panel is in the direction of the dark filament, while the left
        and right spectra are offset by about 30 arcsec (as indicated
        by the coordinates). The resolution is
        30\arcsec and 2.3~\kms. Note the depressed, flat-topped profiles
        toward the filament, flanked by brighter, more smoothly peaked
        profiles. The smooth curve overlaid on each spectrum is the
        parameterized fit. }
         \label{fig:panswfig}
   \end{figure*}

Continuing on from the best signal-to-noise inner disk overview of
Figure~\ref{fig:mp30zoom2fig}, further magnification is required to
discern individual features on a printed page. The three images in
Figs.~\ref{fig:mp30zoomnefig}--\ref{fig:mp30zoomswfig} give some
appreciation of the wealth of detail which can be discerned in the
inner disk of M31. Peak \ion{H}{1} brightness is shown in the left
hand panels of each figure, while velocity coherence VC$_{18.5/2.1}$
(defined in \S\ref{sec:obse}) for the same region is shown on the
right. A striking aspect of these images is the appearance of
filament-like minima as well as positive morphological features, as
seen in peak brightness. While positive emission features are not
surprising when imaging the \ion{H}{1} distribution of an external
galaxy, the filamentary minima are quite novel. A further striking
feature is the correspondence of high velocity coherence with the
majority of these local minima. Apparently the filamentary features
having depressed peak brightness are often accompanied by a
substantial broadening of the underlying spectral profile. This trend
is illustrated in Figs.~\ref{fig:pannefig} and \ref{fig:panswfig} (at
30\arcsec and 2.3\kms\ resolution),
where sequences of spectra are shown which contrast the appearance of
these local minima (central panels) with flanking spectra which are
offset perpendicular to each ``dark'' filament by only about 30 arcsec
(left and right panels). The locations of these cross-cuts are shown
in Figs.~\ref{fig:mp30zoomnefig}, \ref{fig:mp30zoomcfig} and
\ref{fig:mp30zoomswfig}. From these spectra it is apparent that the
line profiles toward such local minima in peak brightness are not only
broad, but have a distinctive flat-topped appearance that is strongly
suggestive of self-opacity in the emission line. The smooth curves
overlaid on each spectrum are the parameterized model-fits to each
spectrum that we describe in detail below in
\S\ref{subsec:glob}. Note that the RMS sensitivity of these spectra,
is that given in Table~\ref{tab:sens} of about 1~K, implying that some
of the fluctuations seen toward the flat-topped peaks are quite significant.
The parameterized fits are indeed consistent with a
significant self-opacity of the \ion{H}{1} emission within this entire
system of filamentary local minima.

   \begin{figure*}
   \centering
   \includegraphics[width=16.5cm]{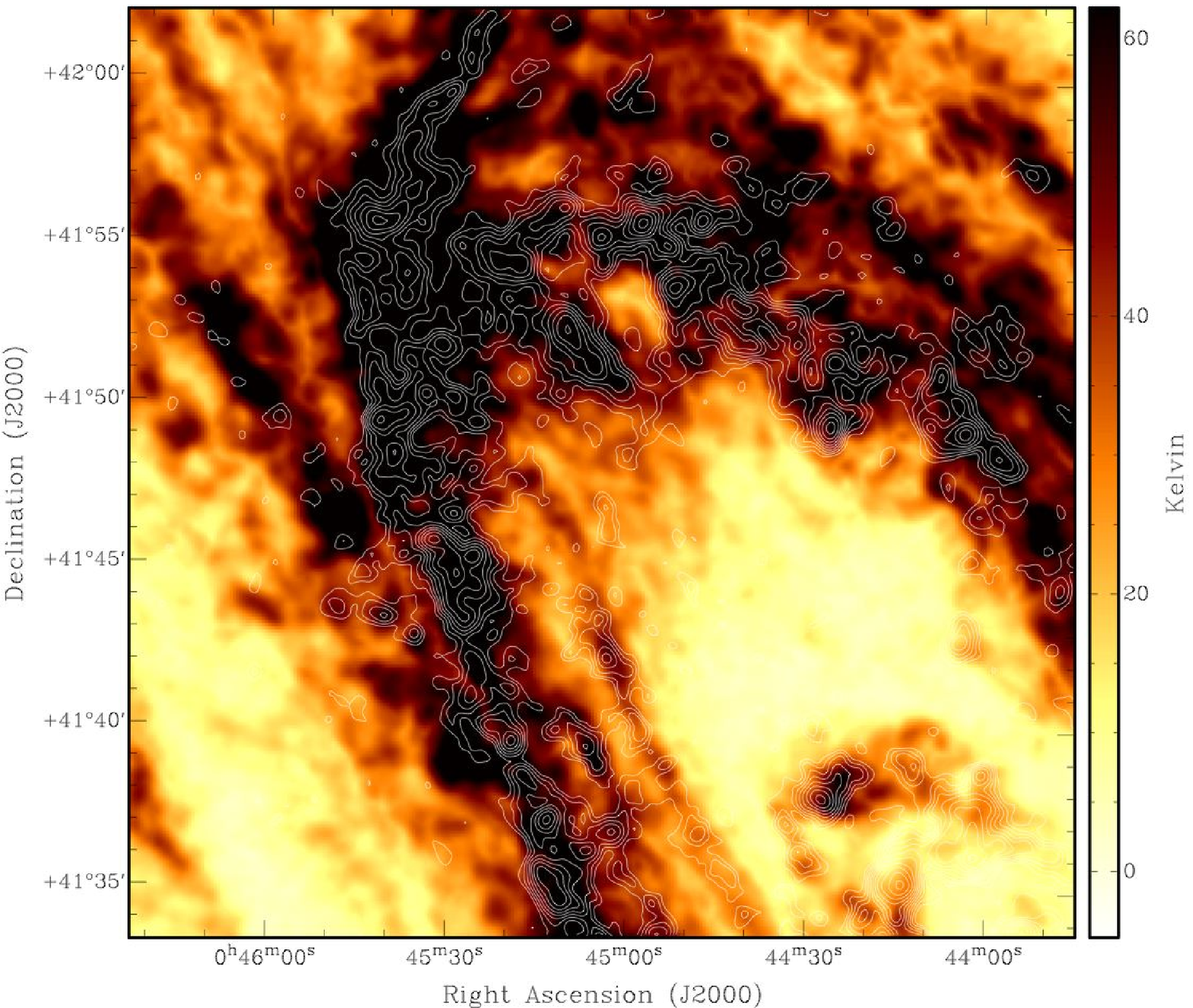}
      \caption{Peak brightness of \ion{H}{1} emission determined at 30
        arcsec and 6 \kms\ resolution of the North-East disk region
        with contours of integrated CO(1-0) (from \citep{niet06})
        superposed at $2.34\times1.2^n$ K-km/s, for n=1,2,3$\dots$.}
         \label{fig:mp30cone}
   \end{figure*}
   \begin{figure*}
   \centering \includegraphics[width=16.5cm]{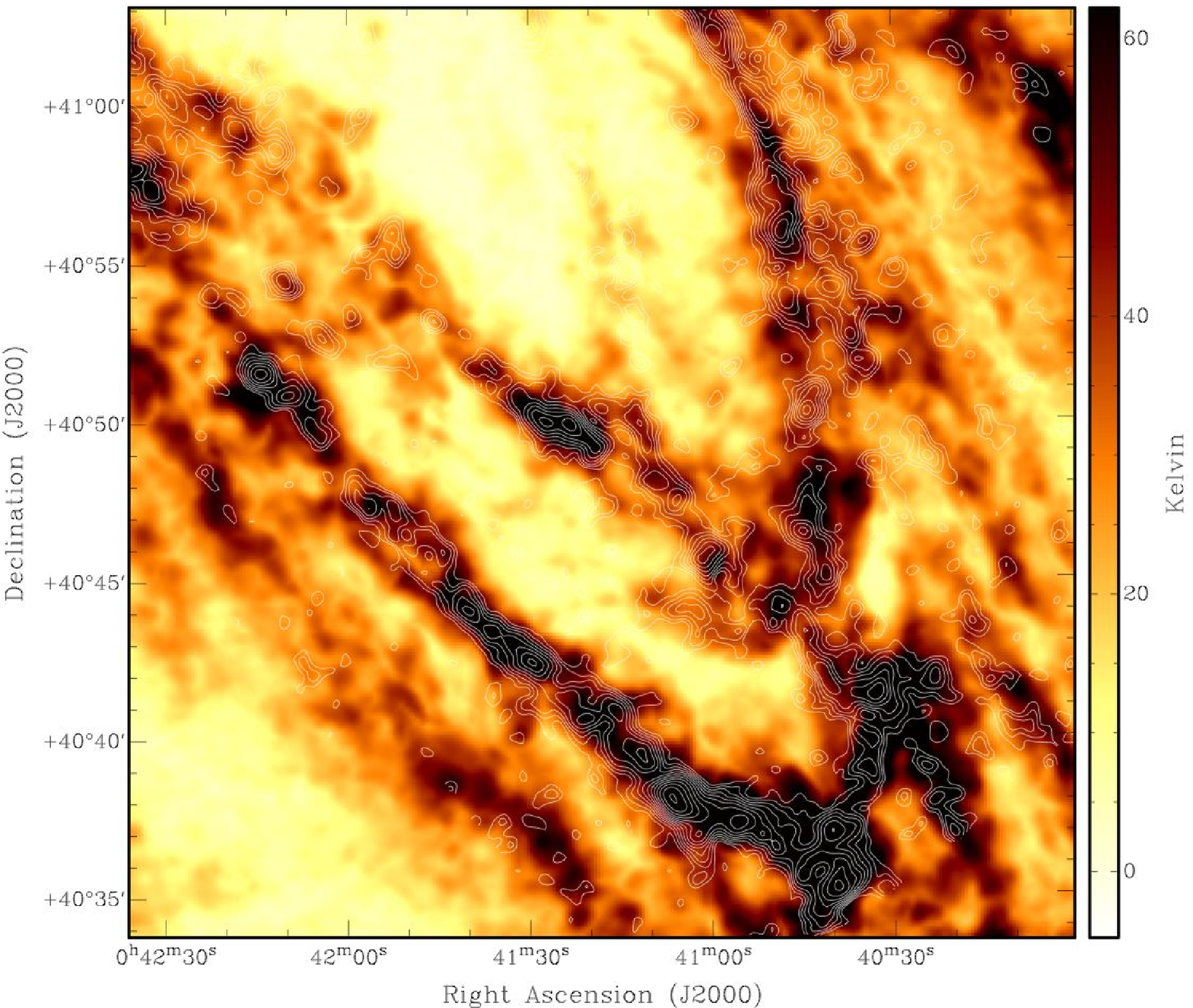}
      \caption{Peak brightness of \ion{H}{1} emission determined at 30
        arcsec and 6 \kms\ resolution of the South-West disk region
        with contours of integrated CO(1-0) (from \citep{niet06})
        superposed at $2.34\times1.2^n$ K-km/s, for n=1,2,3$\dots$.}
         \label{fig:mp30cosw}
   \end{figure*}

\section{Discussion}
\label{sec:disc}

\subsection{Discrete Opaque Features}
\label{subsec:opaq}

The \ion{H}{1} self-absorption (or HISA) phenomenon has been
well-documented in the Galaxy since the pioneering work of
\citet{radh60} and more recently by \citet{gibs05} in a large swath of
the Galactic plane (145$^\circ~>~l~>~75^\circ$). The physical
conditions required to witness HISA are a ``sandwich'' geometry in
which a cooler {\it spin\ } temperature foreground gas is seen toward
a higher {\it brightness\ } temperature background gas at the same
radial velocity. The brightness decrement of a HISA feature is
proportional to this temperature difference in the absence of
additional complicating factors. The simplest circumstance
to imagine is a single warm, semi-opaque background feature which has
a brightness temperature comparable to it's own spin temperature
together with a single cooler, opaque feature in the foreground. In
this simple case the HISA temperature decrement would give some
information about the spin temperature of the foreground
component. More complicated geometries are easily conceivable. Adding
a third warm, semi-opaque foreground component to make a
warm-cool-warm ``sandwich'' would significantly fill-in the detected
temperature decrement. Clearly, the interpretation of HISA temperature
decrements in physical terms is poorly constrained.

The largest individual features seen in the \citet{gibs05} Galactic
HISA sample extend over about 5$^\circ$ at 2~kpc distance, and so have
lengths as large as 175~pc. A subset of the Galactic HISA features,
such as those kinematically associated with the Perseus spiral arm,
are organized over tens of degrees making the complexes at least 1~kpc
long. The departure from the Galactic edge-on geometry to the
$\sim78^\circ$ inclination of M31 \citep{brau91} means that the necessary alignment
conditions for witnessing HISA are essentially eliminated on large
scales in M31. Instead, the ``sandwich'' geometry that can yield
large-scale HISA features in the Galaxy would be projected into
spatially resolved, parallel ``slices'' of semi-opaque gas of
different spin temperature. We identify the filamentary local
minima in peak brightness seen in M31 with the colder opaque features
that are responsible for large-scale HISA in the Galaxy. We use the
term ``self-opaque'' to emphasis the importance of internal optical
depth effects in determining the profile shapes of these features, as
distinct from ``self-absorption'' which implies a substantial
temperature contrast of features that overlap both along the
line-of-sight and in radial velocity.

By comparison to the kpc extent of Galactic HISA complexes, the
self-opaque filamentary minima in M31 (for example the one running
from ($\alpha,\delta)\sim$ (00:45:25,+41:36) to (00:45:55,+41:50) in
Fig.\ref{fig:mp30zoomnefig}) are often in excess of 10~arcmin in
length corresponding to more than 2~kpc. One such linear feature is
seen to cross very near the line-of-sight to the background continuum
source J004218+412926 (=B0039+412), as seen in
Fig.\ref{fig:mp30zoomcfig}. \ion{H}{1} absorption measurements for
this source and several others have been published previously in
\citet{brau92}. We will comment further on these sources below. More
complex local minima of all sizes can be seen in the M31 data wherever
sufficient local intensity is present in both position and
velocity. In contrast, the filamentary network of systematically
broadened profiles (traced by the velocity coherence) are even more
ubiquitous, since profile broadening can be detected independent of
the emission characteristics of the surroundings.

A subset of Galactic HISA features is closely associated with
molecular clouds as traced by OH and CO emission (eg. \citet{gold05}),
while the majority have no apparent CO counterparts
\citep{gibs00}. The conjecture has been made (eg. \citet{gibs05}) that
those HISA features {\it without\ } associated CO detections may
represent an earlier evolutionary phase in molecular cloud formation
and that such features may be organized over kpc-scale regions by the
passage of spiral density wave (or other forms of) shocks. Only after
subsequent cooling and compression would complex molecule-- and
star--formation proceed. Some of the most dramatic self-opaque
features discernible in the M31 peak brightness images appear to lie
on the leading edges of spiral arm structures, for example the feature
extending from ($\alpha,\delta)\sim$(00:45:25,+41:36) to
(00:45:55,+41:50). (We can deduce the upstream edge of bright
\ion{H}{1} features from the counter-clockwise sense of rotation that
follows from the approaching major axis PA~=~232$^\circ$ East of North
and the near-side minor axis PA~=~322$^\circ$). This appears to be
quite a general phenomenon; discernible as a sharp local minimum in
peak \ion{H}{1} along the leading edge of essentially every arm
segment with a suitably bright background. Comparison of the peak
\ion{H}{1} brightness with the integrated CO(1-0) observed with the
IRAM 30m at 23 arcsec resolution \citep{niet06} in
Figs.~\ref{fig:mp30cone}--\ref{fig:mp30cosw} suggests a rather good
correlation of the peak brightness of \ion{H}{1} with integrated CO at
this resolution of about 100~pc at radii less than about 12~kpc; with
a distinct decline of associated CO to larger radii. In contrast, the
filamentary self-opaque minima, particularly those associated with the
leading edges of spiral features, have only very patchy correspondence
with CO(1-0) features, even near 12~kpc.

\subsection{Fitting for Physical Parameters}
\label{subsec:glob}

While some indications for self-opacity in the \ion{H}{1} line are
indicated by qualitative features, like flat-topped spectral profiles,
obtaining quantitative estimates of the ``hidden'' mass in neutral
hydrogen is considerably more challenging. The ``classical'' method of
\ion{H}{1} opacity correction (\citet{schm57}, \citet{hend82},
\citet{brau92}) has consisted of calculating the column density from:
\begin{equation}
N_{HI} = -1.823 \times 10^{18} T_S \Delta V ln \biggl( 1 - {T_B
\over T_S} \biggr) cm^{-2},
\end{equation} 
where a single cool component
spin temperature, T$_S$, is assumed to apply for all
spectra. The relevant spin temperature has been chosen to be
similar to the highest brightness temperature actually observed, for
example T$_S$~=~125~K for the Galaxy \citep{hend82}, or has been
estimated from a fit to the relation between observed absorption
opacity and the emission brightness, T$_S$~=~175~K for M31
\citep{brau92}. In both cases, a single spin temperature is assumed
to apply throughout the galaxy in question, which is clearly a poor
assumption, and the highest observed brightnesses (those with $T_B >
T_S$) have an undefined column density.

Rather than assuming that a single spin temperature might be
appropriate for an entire galaxy, let us consider a more realistic
scenario in which the major atomic clouds in a galactic disk are
approximately isothermal on scales of 100~pc. From \citet{brau97} we
have an estimate of the face-on surface covering factor of such
features within the star forming disk in a sample of nearby galaxies
of about 15\%. This implies that even at the moderately high
inclination of M31, of 78$^\circ$, we are unlikely to have more than a
single major cloud along any line-of-sight in the absence of strong
warping. (Even in the strongly warped case, there need not be
widespread velocity overlap of these components.) In addition, the
good correlation between absorption opacity and the emission
brightness \citep{brau92} in both the Galaxy and M31 suggest that the
contribution of warm (5000 -- 10$^4$~K), optically thin gas in the
extended environment of major clouds to the emission brightness is
only a few Kelvin, although distributed over the relevant line-width
of some 10's of \kms. These considerations suggest that detailed
modeling of the peak (and not the wings) of an emission profile as an
isothermal cloud may provide a useful estimate of the underlying
physical parameters, provided the physical and velocity resolution is
sufficiently high (about 100~pc and 2~\kms) and the galaxy is viewed
at an inclination of less than about 81$^\circ$.

Such an approach has been considered previously by \citet{rohl72}, who have
documented the precision and expected biases that might pertain to
parameter estimation as a function of the peak signal-to-noise for an
{\it unblended,} isothermal cloud. We will return to the question of
the expected uncertainties in such an approach below.

The line profile due to a spatially resolved {\it isothermal\ } \ion{H}{1}
feature in the {\it presence of turbulent broadening\ } can be written as
\begin{equation}
T_B(V) = T_S\{1-exp[-\tau(V)]\}
\label{eqn:tb}
\end{equation}
with
\begin{equation}
\tau(V) = {5.49\times10^{-19}N_{HI} \over
  T_S\sqrt{2\pi\sigma^2}} exp\Biggl(-0.5{V^2 \over
  \sigma^2}\Biggr)
\label{eqn:tau}
\end{equation}
where the velocity dispersion, $\sigma$, has units of \kms\ and is the
quadratic sum of an assumed thermal and nonthermal contribution
$\sigma^2 = \big(\sigma_T^2+\sigma_{NT}^2\big)$ with the thermal
contribution given by $\sigma_T=0.093 \sqrt T_k$, for a kinetic
temperature, $T_k$. While such profiles have a Gaussian shape for low
ratios of the \ion{H}{1} column density (in units of cm$^{-2}$) to
spin temperature, $N_{HI}/T_S$ they become increasingly flat-topped
when this ratio becomes high. For simplicity
we assume that the \ion{H}{1} spin and kinetic temperatures are
equal, $T_k~=~T_S$, which should apply to regions of moderate volume density
although not in general, cf. \citet{lisz01}. We do not consider the
possibility of self-absorption (as discussed above in
\S~\ref{subsec:opaq}) due to multiple temperature components along
individual lines-of-sight, despite seeing occasional evidence for this
phenomenon on small scales. Simple simulations to explore such
multi-temperature models demonstrate the great challenges of
meaningfully constraining the absorber properties. Even negligible
masses of cold \ion{H}{1} can result in dramatic modulation of the
composite spectrum. The additional free parameters associated with
such a composite spectrum make for an intractable fitting problem.

\begin{figure*}
\centering
\includegraphics[width=8.25cm]{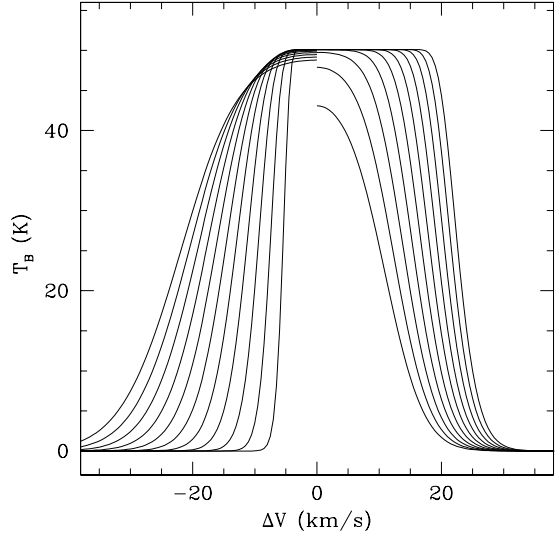}
\caption{ Illustration of model spectra. Shown is a sequence of
``half-'' spectra with increasing $\sigma_{NT}$ = 2 to 12 by 1 \kms\
at a fixed log(N$_{HI}$) = 22 and log(T$_S$) = 1.7 (on the left) as
well as a sequence of increasing log(N$_{HI}$) = 21.5 to 23.5 by 0.2
at a fixed log(T$_S$) = 1.7 and $\sigma_{NT}$ = 7 \kms\ (on the
right).}
\label{fig:mkspcf}
\end{figure*}

We show examples of model isothermal (half-)spectra in
Fig.~\ref{fig:mkspcf}. Shown in the figure are a sequence of
increasing $\sigma_{NT}$ = 2 to 12 by 1 \kms\ at a fixed log(N$_{HI}$)
= 22 and log(T$_S$) = 1.7 (on the left) as well as a sequence of
increasing log(N$_{HI}$) = 21.5 to 23.5 by 0.2 at a fixed log(T$_S$) =
1.7 and $\sigma_{NT}$ = 7 \kms\ (on the right). We found the
best-fitting model spectra for the brightest spectral feature along
each line-of-sight using the 30'' data with full velocity resolution
when the observed peak brightness temperature exceeded 10~K (10 times
the RMS noise) and the truncated peak (where brightnesses greater than
50\% of the peak were encountered) spanned at least five independent
velocity channels. This truncation was done to both isolate single
spectral components from possibly blended features as well as
eliminating potential broad wings from the fit. The data were compared
to a pre-calculated set of model spectra spanning the range
log(N$_{HI}$) = 20.0 to 23.5 by 0.01, log(T$_C$) = 1.2 to 3.2 by 0.01
and log($\sigma_{NT}$) = 0.3 to 1.5 by 0.04. A search in velocity
offset was done over displacements of $-$4 to +4 by 1~\kms\ with
respect to the line centroid as estimated by the first moment of each
truncated peak. We purposely limited the range of log(T$_C$) to that
expected for the Cool Neutral Medium (CNM) where opacity effects might
be expected to occur, but included sufficiently high temperatures (at
least ten times the observed peak brightness temperature) that
negligible opacity solutions are always available for the
fit. Lines-of-sight which might be dominated by the Warm Neutral
Medium (WNM) with spin temperature of perhaps 8000~K, will be fit
by a moderately high T$_S$ (since such profiles would not display
opacity effects), combined with an apparent ``non-thermal'' dispersion
of about 10~\kms. There is no method to discriminate between such
intrinsically warm gas and cooler gas with a comparable non-thermal
broadening of the profile. All that can be discriminated from the
profile shape in the first instance is the presence or absence of
opacity effects. Only if significant opacity effects are detected
(implying $\tau\ge1$) can the physical parameters be
constrained. However, since our primary goal with this procedure is
the correction of apparent column density, this is precisely the
regime in which we can expect some utility of this fitting
method. This conclusion is consistent with the experience of
\citet{rohl72} who indicate that useful parameter fits are only
achieved with a peak signal-to-noise greater than about 10 (such as we
require in our fitting) paired with peak opacities in excess of
unity. Much higher signal-to-noise is required for a meaningful fit at
lower peak opacities. Systematic biases of the fit parameters are
expected to be below about 10\% in this regime.

   \begin{figure*}
   \centering
   \includegraphics[width=16.5cm]{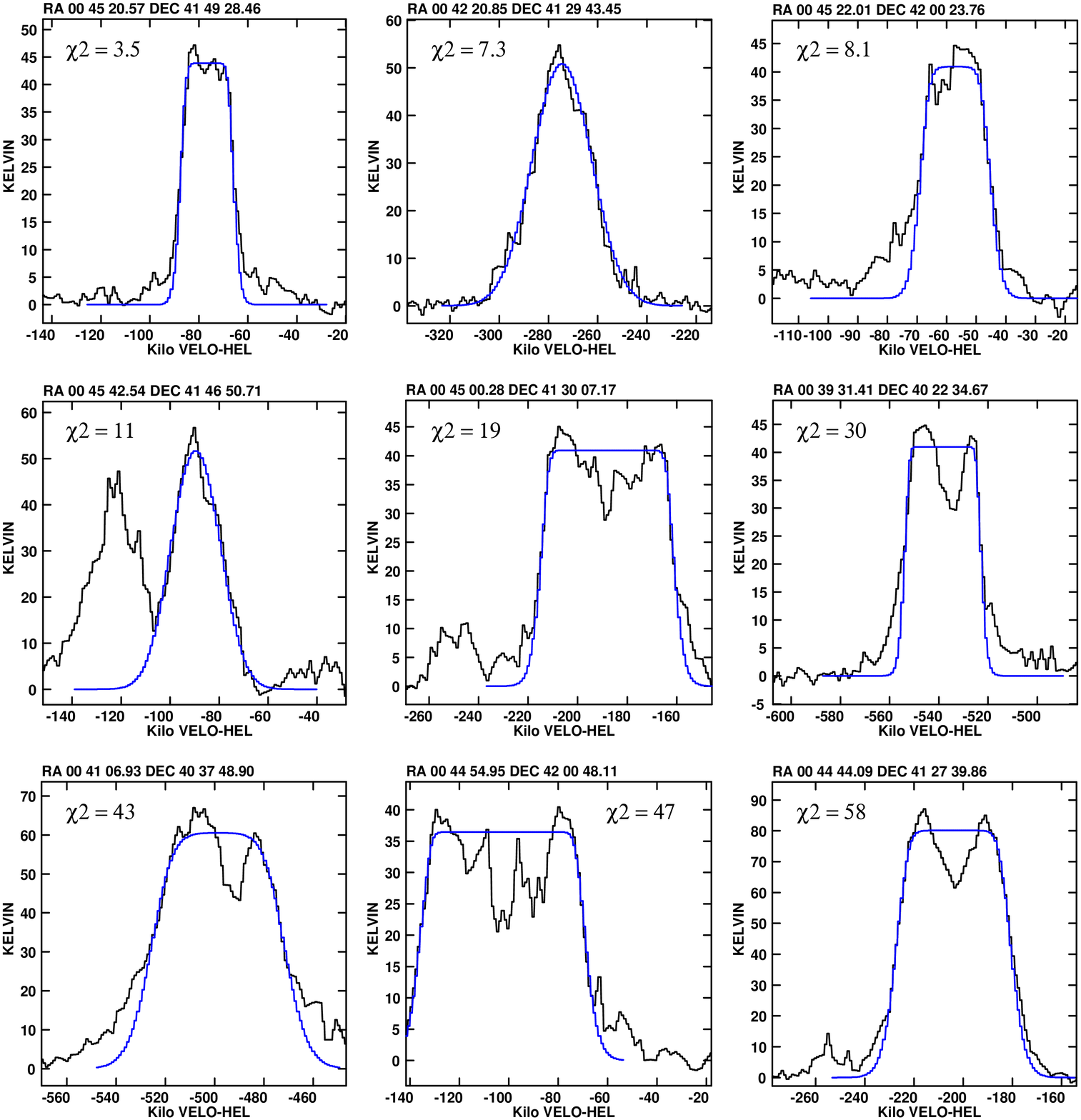}
     \caption{ Illustration of spectral fits. Actual spectra at the
       positions indicated above each box with best-fitting models
       overlaid (smooth curves). The goodness-of-fit is
       indicated for each spectrum with the normalized
       $\chi^2$. Goodness of fit decreases from left-to-right and
       top-to-bottom, primarily due to the effects of velocity
       blending of physically unrelated features. A cut-off $\chi^2
       \sim 25$ was found to best distinguish blended from unblended
       spectra. }
         \label{fig:spcfig}
   \end{figure*}

Plausible spectral fits were possible for most
lines-of-sight. However, there are locations in the M31 disk where
emission features which are likely at significantly different distances
along the line-of-sight become strongly blended in velocity. Such
circumstances are most apparent in position-velocity images made
parallel to the major axis, in which continuous spiral arm segments
can be tracked as linear features with approximately constant
orientation (as documented by \citet{brau91}). The gradient of
velocity with major axis distance,
\begin{equation}
 {dV \over dX} = {V_c(R) sin(i(R)) \over R }
\end{equation}
is proportional to
the local projected rotation velocity divided by the radial distance
of the feature within the galaxy. The ocurrence of apparently
intersecting linear features in position-velocity plots is a strong
indication for velocity blending of physically unrelated features
which reside at different radii in the galaxy. The most prominent
instances of such velocity overlap occur in four locations within M31
at major axis distances of both $\pm30$ arcmin and minor axis
distances of both $\pm10$ arcmin.

The challenges of velocity blending are illustrated by the sequence of
individual spectra with overlaid fits shown in
Fig.~\ref{fig:spcfig}. The sequence of spectra from left to right and
top to bottom are ordered by decreasing quality of fit, as measured by
the normalized $\chi^2 = \Sigma_i (D_i-M_i)^2/(N\sigma^2)$. Our
fitting method could deal successfully only with instances of
sufficiently separated spectral features, where the local minima drop
below 50\% of the peak. More blended features are straddled with a
single broad profile fit. From a careful examination of individual
cases of blended and unblended profiles, based on the presence or
absence of intersecting linear position-velocity features as described
above, we established a cut-off in fit quality of $\chi^2 = 25$ that
most effectively seperated these two regimes. Only those fits with a
$\chi^2$ below this cut-off were retained, resulting in rejection of
some 4\% of the total. Concentrations of rejected fits are located in
the expected regions (noted above). 

\begin{figure*}
   \centering
   \includegraphics[width=16.5cm]{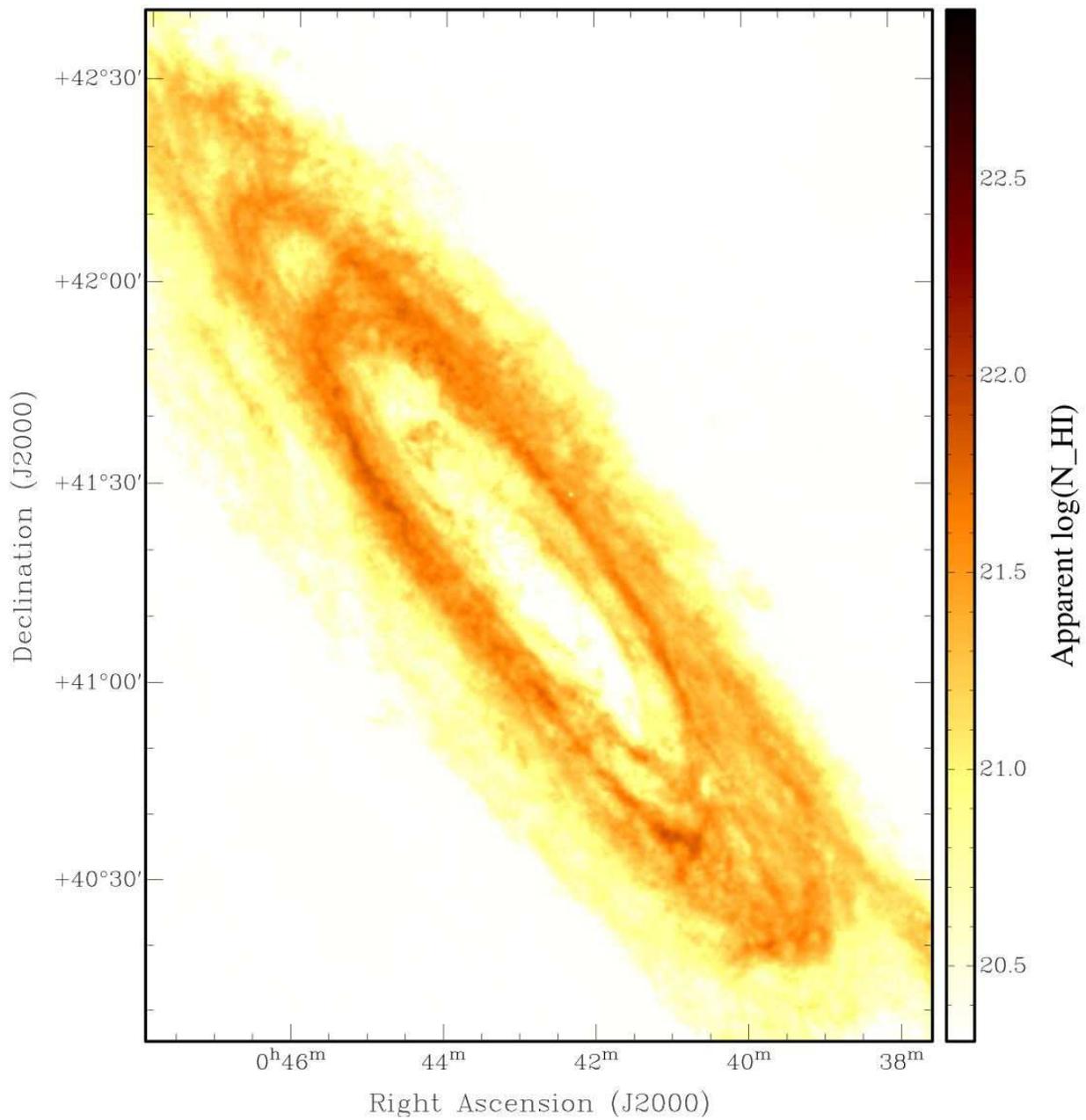}
      \caption{Column density of \ion{H}{1} assuming negligible
	opacity of the central 25\% of the survey region. }
         \label{fig:n30lnhvfig}
   \end{figure*}

\begin{figure*}
   \centering
   \includegraphics[width=16.5cm]{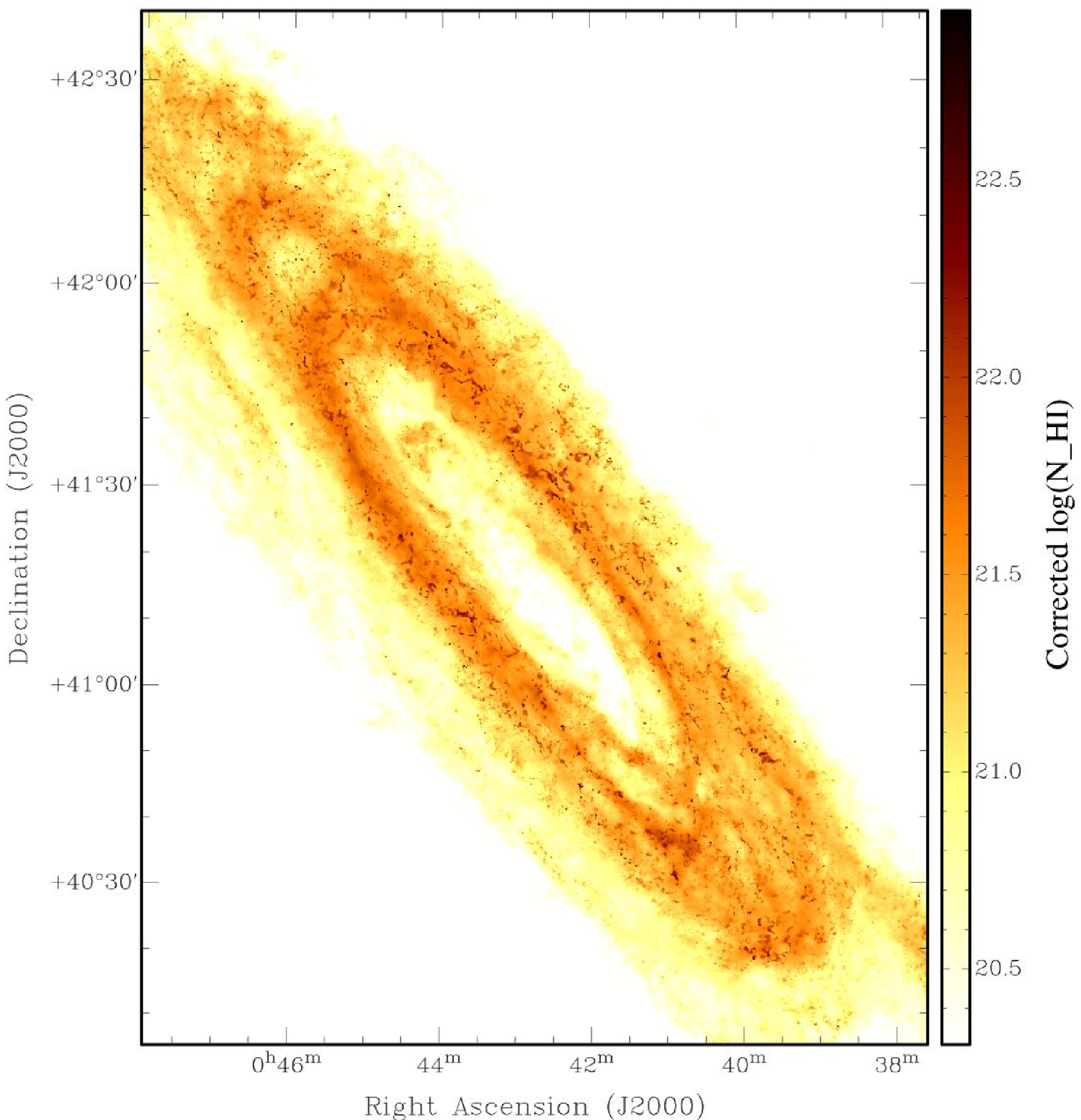}
      \caption{Column density of \ion{H}{1} derived from a fit to each
	spectrum of the isothermal model described in
	\S\ref{subsec:glob} for the central 25\% of the survey
	region. }
         \label{fig:n30lnhcfig}
   \end{figure*}

The results of our spectral fitting were recorded as images of the
physical parameters, $N_{HI}^{Fit}, T_S$ and $\sigma_{NT}$ together
with the integral of brightness temperature over velocity that
corresponds to the column density of the fit, $\int T_B^{Fit}dV$. This
allowed calculation of a total corrected column density image from,
\begin{equation}
N_{HI}^{Tot} = N_{HI}^{Fit} +
1.823\times10^{18}\biggl(\int T_B^{Tot}dV-\int T_B^{Fit}dV\biggr)
\end{equation}
where $\int T_B^{Tot}dV$ is just the usual image of integrated
emission. In this way any spectral component not accounted for by the
fit to the peak was retained, albeit with the assumption of negligible
opacity for this residual. Those lines-of-sight with insufficient fit
quality ($\chi^2 > 25$) were assumed to have negligible opacity.  The
two column density images (assuming negligible optical depth and
fitting for it) are contrasted in Figs.~\ref{fig:n30lnhvfig} and
\ref{fig:n30lnhcfig}. Substantially more fine-scale structure is
apparent in the opacity-corrected column density, primarily in the
form of compact features and narrow filaments. The narrow filaments,
in particular, correspond to those seen in the images of velocity
coherence, shown in
Figs.~\ref{fig:mp30zoomnefig}--\ref{fig:mp30zoomswfig}.  Peak column
densities are enhanced by about an order of magnitude over the
uncorrected version. 

Given the uncertainties in the fit parameters, we do not attach undue
significance to each pixel in the images of corrected column density,
since they may easily be in error by a factor of 2 or more, but are
encouraged by the spatial continuity of the solutions, since the fit to
every line-of-sight is carried out independently of all others. Some
indication for the relevance of the fit results can be obtained by
comparison with high signal-to-noise observations of \ion{H}{1}
absorption toward background continuum sources. As noted previously,
several bright background continuum sources that have been studied by
\citet{brau92} fall fortuitously near regions of significant
self-opacity, particularly the sources B0039+412 and B0044+419 for
which peak optical depths of $\tau_{max}=0.46\pm0.02$ and
$\tau_{max}=1.06\pm0.06$ are observed. (Note that $(1-e^{-\tau})$
rather than $\tau$ is
tabulated in the reference.) The model fits show very
substantial structure in the vicinity of these background sources,
making it difficult to estimate the likely line-of-sight parameters. We have
determined the average peak model opacity over a 3$\times$3 pixel
region that is offset azimuthally from the direction of each of these
sources by one beamwidth. For these two lines-of-sight this yields
$\tau_{max}=0.4\pm0.4$ and $\tau_{max}=1.1\pm0.8$. Although the
pixel-to-pixel variation is very large, and may well represent real
structure rather than fitting errors, the order-of-magnitude of the
opacity is estimated correctly in these cases.

The total detected \ion{H}{1} mass of M31 is increased by the opacity
corrections by about 30\%, from 5.76$\times 10^9$ to 7.33$\times 10^9$
M$_\odot$. In view of the necessity of discarding all blended
profiles, this should be regarded as a lower limit to the actual mass
correction (within the limitations of our approach).

\begin{figure*}
   \centering
   \includegraphics[width=16.5cm]{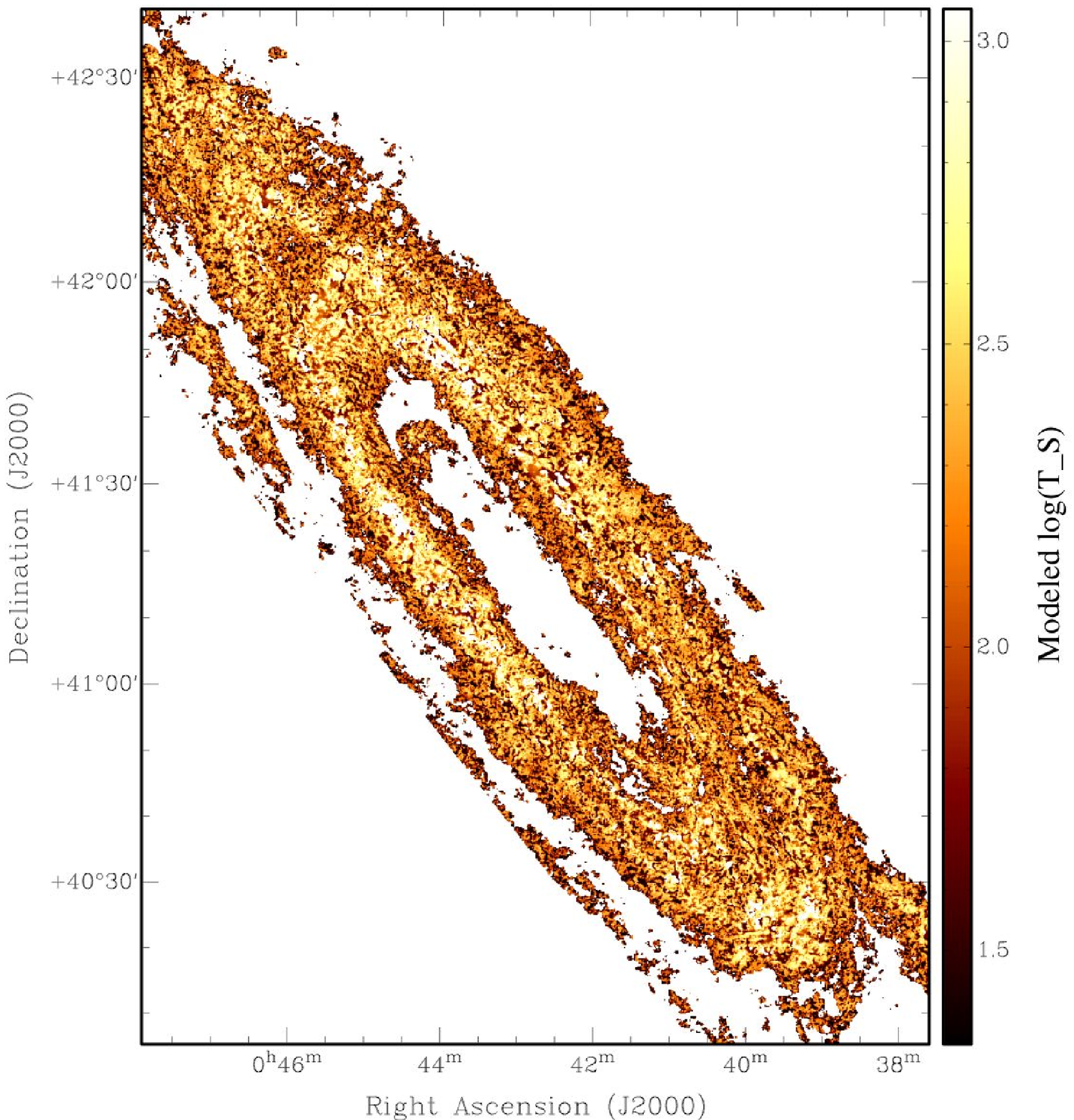}
      \caption{The logarithm of the spin temperature of \ion{H}{1}
	derived from a fit to each spectrum of the isothermal model
	described in \S\ref{subsec:glob} for the central 25\% of the
	survey region. }
         \label{fig:n30ltcfig}
   \end{figure*}

\begin{figure*}
   \centering
   \includegraphics[width=16.5cm]{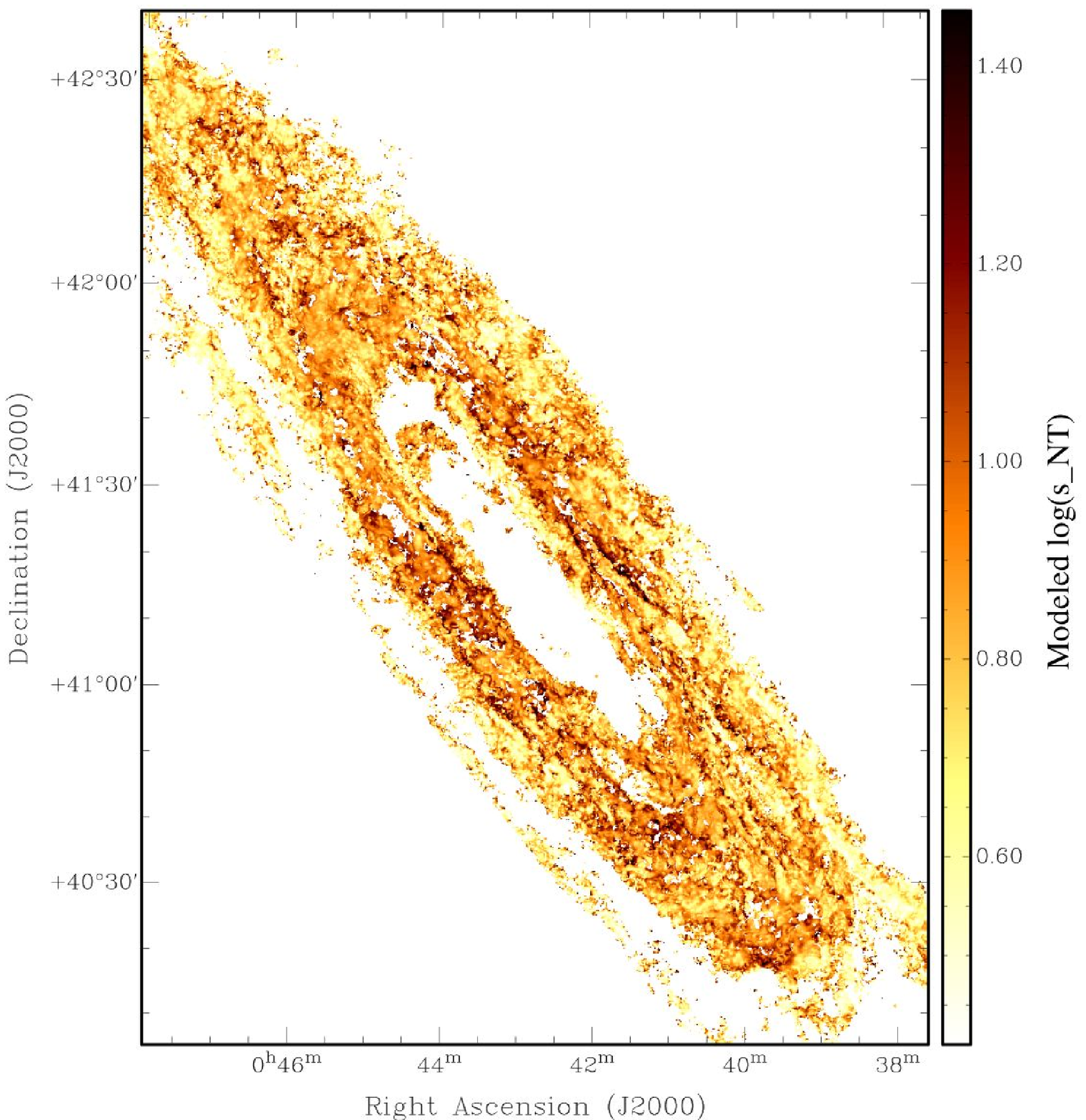}
      \caption{The logarithm of the non-thermal velocity dispersion of
	\ion{H}{1} derived from a fit to each spectrum of the
	isothermal model described in \S\ref{subsec:glob} for the
	central 25\% of the survey region. }
         \label{fig:n30sntfig}
   \end{figure*}

The best-fit spin temperature and non-thermal velocity dispersion
are shown in Figs.~\ref{fig:n30ltcfig}--\ref{fig:n30sntfig}. Regions
of significant opacity (log($N_{HI}$)~$>$~22) have spin temperatures
of less than about 100~K and are often organized into patchy
filamentary complexes. While there is some organized spatial structure
in the distribution of non-thermal velocity dispersion, it is not
obviously correlated with either the total column density or spin
temperature. Not surprisingly, there is a significant correlation of
the non-thermal velocity dispersion of the fits with the image of
velocity coherence (shown in the right hand panels of
Figs.~\ref{fig:mp30zoomnefig}--\ref{fig:mp30zoomswfig}). 

\subsection{Radial Trends}
\label{subsec:radt}

The basic radial trends of observed and derived quantities are
illustrated in Fig.~\ref{fig:ronlyfig}. The peak observed brightness
temperature (Panel~A) is seen to first increase dramatically with
radius out to about 12~kpc and then subsequently to decline out to the
the largest detected radii. Even the median peak brightness exceeds
50~K near 12~kpc, while the distribution extends out beyond
100~K. This can be compared with the very general trend documented by
\citet{brau97} for a positive gradient in peak brightness to occur
within the actively star forming disk of nearby galaxies. The
unprecented physical sensitivity obtained for M31 allows us to track
the atomic gas properties out to more than three times larger radii.
The apparent column density distribution in Panel~B follows a similar
trend, illustrating the exponential character of both the increase to-
and the subsequent decline from- the peak near 12~kpc. A 6~kpc
scale-length exponential function is overlaid on the plots in Panels
B, C, D and H to illustrate how well this scale-length describes the
gas mass density. Intriguingly, the old stellar population of M31, as
traced by 3.6 $\mu$m Spitzer imaging between 1 and 12~kpc radius is
best-fit with a 6.08$\pm$0.09~kpc scale-length exponential
\citep{barm06}. The corrected \ion{H}{1} column density, shown in
Panel~C, has a median ridge-line that closely tracks the apparent
column, but also demonstrates that outliers to higher corrected column
occur primarily at radii between about 7 and 30~kpc. Adding in the
molecular contribution to mass surface density (as described in detail
in \S~\ref{subsec:sfrd}) in Panel~D yields only minor changes to the
radial trend of the atomic column density. The most significant is a
slight addition of mass density near 12~kpc, where the molecular
contribution permits the total to better track the 6~kpc exponential
to slightly smaller radii. Inside of about 12~kpc is a sharp cut-off
that can be described with a 3~kpc exponential scale-length. Beyond
about 28~kpc, where opaque \ion{H}{1} is no longer found, there is a
significant truncation of the disk exponential (neutral) mass density,
corresponding to a 2~kpc scale-length. The corresponding face-on
column density (scaled by 1/cos(78$^\circ$)) where this outer
truncation sets in, is 5$\times 10^{19}$cm$^{-2}$.

The radial distribution of velocity coherence (Panel~E) shows a
similar trend, rising rapidly out to about 10~kpc, then declining out
to about 30~kpc and then plummeting. Both the derived non-thermal
velocity dispersion (Panel~F) and spin temperature (Panel~G) show
some broad systematic decline with radius beyond 10~kpc. For the
spin temperature, it is only the lower envelope of the distribution
which represents ``opaque'' fits. These ``opaque'' cool component
temperatures are as low as 20~K near 7 and 25~kpc, but increase to
about 60~K near 10~kpc. This is quite distinct from the most highly
populated ridge-line, which represents predominantly ``transparent''
lines-of-sight. The ridge-line shows a similar trend, but is offset to
higher temperatures by about an order of magnitude. Current massive
SFR density (as defined below in \S~\ref{subsec:sfrd}) is
shown in Panel~H. The apparent nuclear peak in SFR density is due to other
processes (as discussed below), while current coverage only extends to
about 20~kpc. Current SFR density is highly peaked near
12~kpc, where gas mass densities are highest and does not track the
6~kpc exponential of the median gass mass density.

\subsection{The Surface Density of Gas Mass and Star Formation Rate}
\label{subsec:sfrd}

A particular form of correlation of observable parameters deserves
special consideration, namely that between the surface density of gas
mass and the star formation rate. Many authors have addressed the
relationship between these properties on both a global and local
level, beginning with the pioneering study of \citet{schm59} and most
recently by \citet{kenn07} and \citet{thil07}, who probe the resolved
relationship of these parameters down to scales of 500~pc within
NGC~5194 (M51a) and 400~pc within NGC~7331 respectively. Given the
proximity of M31 we can extend this work by more than an order of
magnitude to smaller surface area at spatial scales of 100~pc, albeit
with a substantial inclination correction that introduces some
potential linear smearing (by a factor of 4.8) in one dimension. Mass
and star formation rates have been calculated after background
subtraction and spatial smoothing of all relevant images to the same
30~arcsec (113~pc) Gaussian resolution as applies to the \ion{H}{1}
properties.

The mass surface density was calculated from the integrated CO(1--0)
emission observed with the IRAM telescope by \citet{niet06} assuming a
mean atomic mass of 1.36 m$_H$ and an N$_{H_2}$/I$_{CO}$ conversion
factor of 2.8$\times 10^{20}$cm$^{-2}$/K-\kms. This differs from the
procedure adopted by \citet{kenn07}, (hereafter K07)
who keep the comparison in terms of hydrogen nuclei (rather than
including a Helium mass correction). The helium mass correction
amounts to a shift by +0.13 dex in mass density.  Note that the CO
data only extend over the central 2$\times$0.5 degrees (or
28$\times$32 deprojected kpc in diameter). Radii beyond about
15--20~kpc therefore only have an \ion{H}{1} contribution to the
calculated mass density. Although the CO brightness at the edges of
the sampled region is generally faint and has a low surface covering
factor, isolated pockets of CO may well be present at larger radii.

The star formation density was calculated following the methodology of
\citet{thil07} from the Spitzer IRAC 8$\mu$m and MIPS 24$\mu$m surface
brightness of \citet{gord06} together with the GALEX FUV image of
\citet{thil05} and \citet{gild07}. The ``bolometric'' luminosity was
assumed to be, $L(bol)=\nu_{FUV}L_{\nu ,obs}(FUV)+(1-\eta)L(IR)$ after
correcting the FUV image for foreground extinction by the Galaxy, and
assuming $\eta$=0.32. The term ``bolometric'' in this context is a
misnomer, since the quantity is not assumed to represent a true
integral of the entire SED, but merely to provide sensitivity to both
directly emitted UV and re-radiated IR light. The value of $\eta$,
which represents the fraction of the IR luminosity due to an older
stellar population, is expected to vary substantially with environment
within a galaxy. Ideally, this would also be modeled in a position
dependent way, but that is beyond the scope of the current discussion.

The total (3-1100um) IR luminosity per pixel was estimated
using the IRAC/MIPS data and the relation,
\begin{equation}
log\bigg[{L(IR) \over \nu L_\nu(24\mu m)} \bigg] = 1.06 + 0.475
\ log\bigg[{F_\nu(8\mu m) \over F_\nu(24\mu m)} \bigg]
\end{equation}
and then converted to a star formation rate density using the
projected pixel area and the SFR calibration: 
\begin{equation}
 log\ SFR(bol) (M_\odot yr^{-1}) = log\ L(bol) (L_\odot) - 9.75.
\end{equation}
The 24$\mu$m data only extend over the brighter disk region of 
$\sim$3$\times$1 degrees. 

\begin{figure*}
   \centering
   \includegraphics[width=16.5cm]{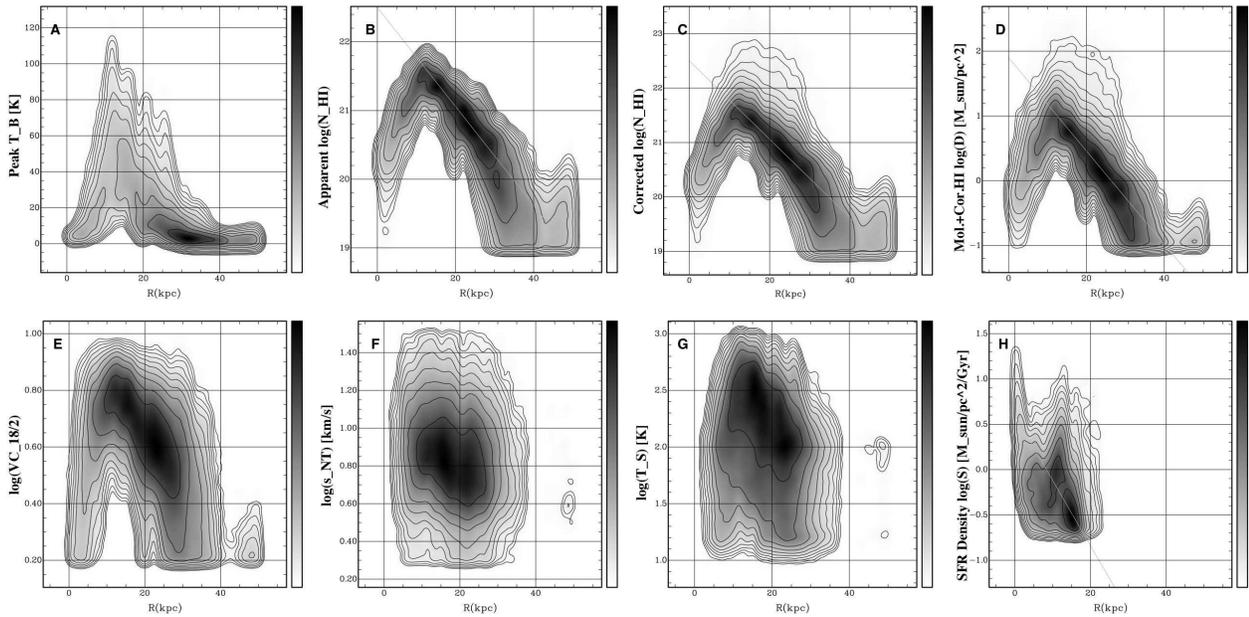}
      \caption{The radial distribution of several diagnostic
        properties of M31. The plotted quantities are the peak
        \ion{H}{1} brightness temperature, apparent \ion{H}{1} column
        density, corrected \ion{H}{1} column density, the face-on gas
        mass surface density in M$_\odot$pc$^{-2}$, \ion{H}{1}
        velocity coherence, the best-fit non-thermal velocity
        dispersion (labelled log(s\_NT)) the best-fit cool \ion{H}{1}
        temperature (labelled log(T\_S)) and the estimated face-on
        star formation rate density in M$_\odot$pc$^{-2}$Gyr$^{-1}$
        based on FIR and UV surface brightness. Note that the central
        peak in panel H is not due to star formation (see
        text). Contours are drawn at factors of two in correlation
        density beginning at 0.1\% and ending at 51\%. The greyscale
        extends from 0 and 100\% with a square-root transfer
        function. The gray solid line in panels B, C, D and H is an
        exponential with 6.0 kpc scale-length.}
         \label{fig:ronlyfig}
   \end{figure*}

Only pixels with significant emission in the spatially smoothed images
were used for the comparison, which resulted in thresholds at a
brightness of 0.7 K-\kms\ for the CO emission, \ion{H}{1} columns
greater than 19 in the log and sky-subtracted IR brightness greater
than 0.06 MJy/sr at 24$\mu$m and 0.1 MJy/sr at 8$\mu$m. As can be seen
in Fig.~\ref{fig:ronlyfig}, the spatial coverage and sensitivity in
the IR only extends out to a radius of about 20~kpc.

\begin{figure*}
   \centering
   \includegraphics[width=16.5cm]{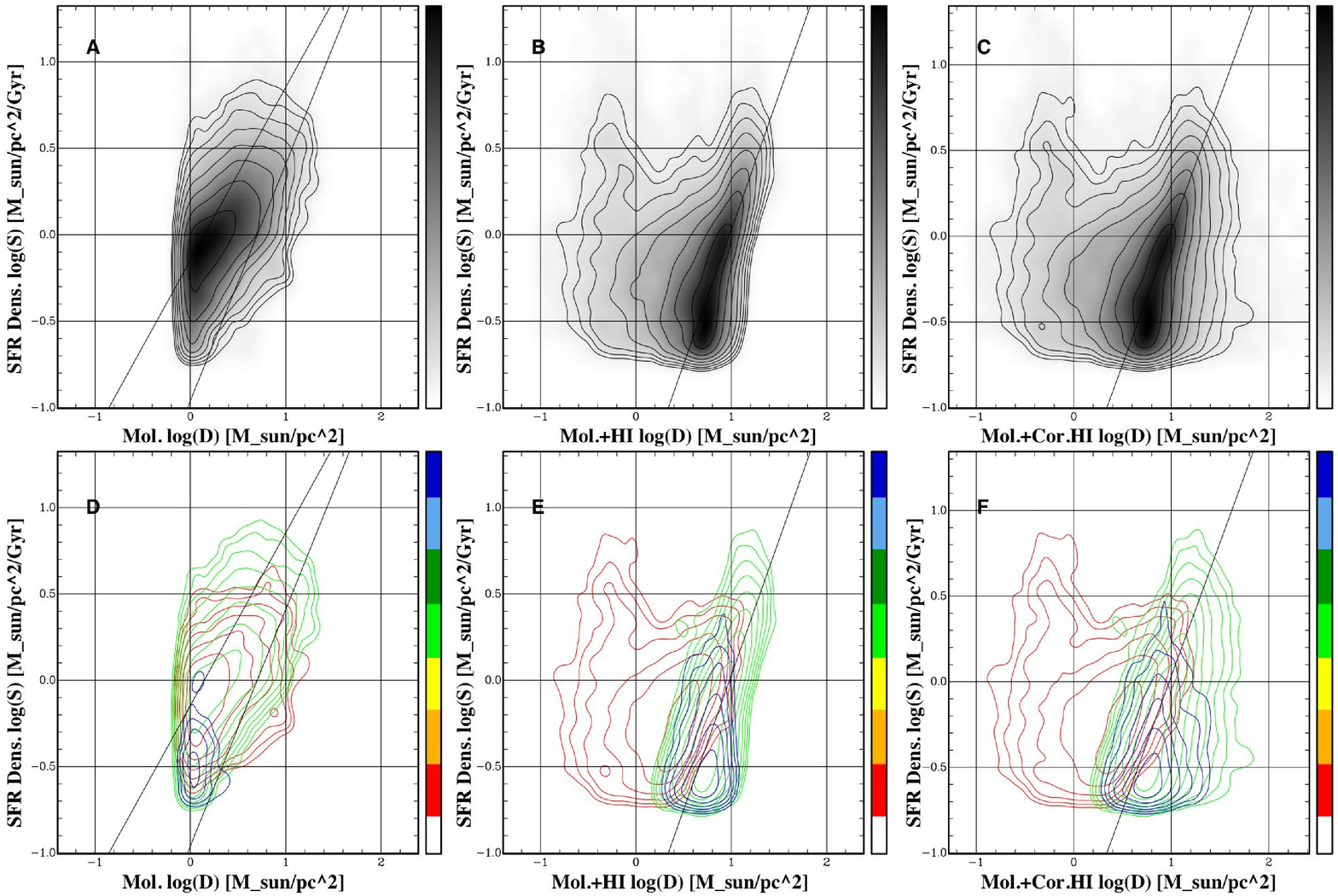}
      \caption{Star formation rate density versus gas mass density in
      M31. The face-on gas mass surface density has units of M$_\odot$pc$^{-2}$
      and the face-on star formation rate density,
      M$_\odot$pc$^{-2}$Gyr$^{-1}$. The mass
      surface density is calculated with molecular only--, molecular
      plus apparent \ion{H}{1}-- and molecular plus opacity-corrected
      \ion{H}{1} column density from left to right. The data at all
      radii are shown on the top, while the radially binned data is
      shown on the bottom. Contours are drawn at factors of two in
      correlation density beginning at 0.5\% and ending at 64\%. The
      greyscale extends from 0 and 100\% with a square-root transfer
      function. The colors red, green and blue correspond to bins in
      galactocentric radius of 0--8, 8--16 and 16--24 kpc. The
      diagonal lines of slope 1.37 (molecular only) and slope
      1.56 (molecular plus atomic) are the relations given in
      \citet{kenn07}. A second line of unit slope is
      plotted in the molecular-only plots, corresponding to a gas
      depletion timescale of 1~Gyr. } 
         \label{fig:sfrfig}
   \end{figure*}

We contrast three different versions of the star formation law in
Fig.~\ref{fig:sfrfig} utilizing the molecular only, the molecular plus
apparent \ion{H}{1} and the molecular plus opacity-corrected
\ion{H}{1} column density. The molecular-only plots are overplotted
both the relation of K07 with slope 1.37, a value which has also been
found by \citet{heye04} in M33, as well as a line with unit slope
corresponding to a constant gas depletion time-scale, as advocated by
\citet{wong02}, (hereafter WB02). The slope of the molecular-only
relationship is approximately matched by the simple linear
dependency. The gas depletion time-scale of the plotted line is 1~Gyr
(for a N$_{H_2}$/I$_{CO}$ conversion factor of 2$\times
10^{20}$cm$^{-2}$/K-\kms\ as assumed by Wong \& Blitz). This is
similar to many of the galaxies analyzed in WB02, although occurring
at mass and star formation densities that are at least an order of
magnitude lower within M31. The M51a molecular-only relation of K07 is
both steeper and shifted to higher mass densities by at least 0.3 dex.

The total-gas mass density plots are over-plotted with the relation
fit by K07 to the M51a data with slope 1.56. Note that we have plotted
star formation rate density in units of M$_\odot$pc$^{-2}$Gyr$^{-1}$ rather
than the M$_\odot$kpc$^{-2}$yr$^{-1}$ of K07 implying a shift of 3
dex. The different units of both axes (relative to K07) have been
taken into account for the overplotted curves. Remarkably, the M51a
relation provides a very good fit to the M31 data, despite the fact
that it has been derived in the strongly molecular-dominated inner
disk of M51 and is being applied thoughout the atomic-dominated disk
of M31. The highest contours of correlation probability are about
twice as narrow for the total-gas relation compared to the
molecular-only one. In the relation using apparent \ion{H}{1}, there
is a very steep truncation to mass densities exceeding about
10~M$_\odot$pc$^{-2}$, together with a saturation that sets in near
16~M$_\odot$pc$^{-2}$. This truncation disappears in the
opacity-corrected \ion{H}{1} relation, together with a small shift of
the peak correlation ridge-line to higher mass densities. The
saturation effect near 16~M$_\odot$pc$^{-2}$, is also alleviated
although not entirely eliminated.

Data at the smallest radii ($<$5~kpc) yield the diffuse region of
elevated {\it apparent\ } star formation in the total gas panels of the
figure. The enhanced 24$\mu$m emission from the nuclear regions of M31
is accompanied by both recombination- and forbidden emission line gas
(as first documented by \citet{jaco85}) with line ratios
suggestive of shock heating, and as such is not due to massive star
formation. Essentially no CO is detected from the nuclear region, so
this feature is not apparent in Panels A and D. Beyond about 5~kpc,
the same basic correlation is seen to apply at all radii, with only
the range of the observables showing some systematic variation, but
not the basic ridge-line of their correlation. The sharp truncation of
the molecular-only mass density near 1~M$_\odot$pc$^{-2}$ is a
consequence of the limited sensitivity. The same is true for star
formation rate densities below about 0.2~M$_\odot$pc$^{-2}$Gyr$^{-1}$.

At star formation rate densities below about
0.4~M$_\odot$pc$^{-2}$Gyr$^{-1}$ and total gas surface densities of
about 5~M$_\odot$pc$^{-2}$ there appears to be a departure from the
power-law relationship. The same feature can be seen in the NGC~7331
data of \citet{thil07} at a similar location, although this is 
near the noise floor in both cases. This may indeed be a distinct
threshold for the formation of molecular clouds from which stars will
form. The most substantial contribution to this possible threshold
comes from radii in excess of 15~kpc.

Perhaps the most surprising aspect of Fig.~\ref{fig:sfrfig} is the
good statistical correlation of total gas mass and star formation
density down to spatial scales of about 100~pc and gas masses of only
5$\cdot 10^4$~M$_\odot$, well inside the regime of individual, giant
molecular clouds.



\section{Conclusions}
\label{sec:conc}

Our high resolution and signal-to-noise study of the neutral ISM of
M31 has permitted a number of novel results:

\begin{enumerate}

\item We detect ubiquitous self-opaque \ion{H}{1} features,
  discernible in the first instance as filamentary local minima in
  images of the peak \ion{H}{1} brightness temperature. Such local
  minima are often accompanied by systematically broader line
  profiles, apparent in images of the velocity coherence. High opacity
  features are organized into complexes of more than kpc length and
  are particularly associated with the leading edge of spiral arms.
  We suggest that such features are the counterpart of large-scale
  \ion{H}{1} self-absorption (HISA) features seen in the Galaxy. They
  are the consequence of a ``sandwich'' geometry of two warm,
  semi-opaque layers which flank a colder, opaque core. Such a
  geometry leads to widespread HISA in the edge-on Galactic disk, but
  is spatially resolved into distinct parallel filaments in the
  inclined disk of M31. Just as in the Galaxy, there is only patchy
  correspondence of self-opaque/HISA features with CO(1-0) emission.

\item We have produced images of the best-fit physical parameters;
  spin temperature, opacity-corrected column density and non-thermal
  velocity dispersion, for the brightest spectral feature along each
  line-of-sight in the M31 disk. This represents a major step forward
  in the quantitative assessment of \ion{H}{1} opacity in a galactic
  context. Opaque atomic gas is organized into filamentary complexes
  and isolated clouds down to the resolution limit of 100~pc. The spin
  temperature of opaque regions first increases systematically with
  radius from about 20 to 60~K (for radii of 5 to 12~kpc) and then
  smoothly declines again to 20~K by about 25~kpc. Opacity corrections
  to the column density exceed an order of magnitude in many cases and
  add globally to a 30\% increase in the atomic gas mass over that
  inferred from the integrated brightness under the usual assumption
  of negligible self-opacity. It will be particularly interesting to
  undertake a similar analysis of more nearly face-on systems like M33
  and the LMC, where spectral blending will be even less of an issue.


\item We have constructed the radial distribution of gas mass density
  which illustrates a remarkably extended exponential decline with
  6~kpc scale-length between 12 and 28~kpc. Inside of 12~kpc there is
  a rapid decline in neutral mass density with 3~kpc scale-length,
  while beyond 28~kpc at a face-on column density of 5$\times
  10^{19}$cm$^{-2}$ is a truncation of the atomic disk with 2~kpc
  scale-length. Perhaps fortuitously, the extended disk scale-length
  of 6~kpc matches that of the old stellar population traced by
  3.6~$\mu$m emission.

\item We have extended the resolved correlation of star-formation-rate-
  with gas-mass-density down to the smallest physical scales yet
  reached, of only 100~pc and $10^4$~M$_\odot$; more than an order of
  magnitude smaller in area and mass than has been possible
  previously. Unlike other galaxies, in which the gas mass is
  dominated by molecular gas as traced by CO emission at small radii,
  M31 is dominated by atomic gas at all radii. The relation of
  molecular-mass- to star-formation-density has a large dispersion,
  but has a slope near unity and a corresponding gas depletion
  timescale of about 1~Gyr, similar to many galaxies studied by
  \citet{wong02}. The relation between total-gas-mass- and
  star-formation-rate-density is significantly tighter and is fully
  consistent {\it in both slope and normalization\ } with the relation
  found in the molecule-dominated disk of M51 by \citet{kenn07} at
  500~pc resolution. Together, the M31 and M51 data demonstrate the
  same relationship spanning mass-densities from about 5 to 700
  M$_\odot$pc$^{-2}$. Use of opacity-corrected \ion{H}{1} columns
  yields a more symmetric distribution with less evidence of
  truncation effects than with the apparent \ion{H}{1} column. Below
  about 5 M$_\odot$pc$^{-2}$, there is a down-turn in
  star-formation-density which may represent a real local threshold
  for massive star formation. The corresponding threshold cloud mass
  is about 5$\cdot 10^4$~M$_\odot$.

\end{enumerate}



\acknowledgments

We acknowledge the very useful comments of an anonymous referee. We
are grateful to John Romein for his contributions to Miriad code
parallelization and enabling grid processing on the Netherlands
Grid. The Westerbork Synthesis Radio Telescope is operated by ASTRON
(Netherlands Foundation for Research in Astronomy) with support from
the Netherlands Foundation for Scientific Research (NWO). The National
Radio Astronomy Observatory, which operates the Green Bank Telescope,
is a facility of the National Science Foundation operated under
cooperative agreement by Associated Universities, Inc.



{\it Facilities:} \facility{WSRT}, \facility{GBT}.

\clearpage

\end{document}